\documentclass[11pt,a4paper]{article}

\usepackage{jinstpub} 

\usepackage{siunitx}
\usepackage{pdfpages}
\usepackage{float}
\usepackage{xcolor}
\usepackage{ulem}
\usepackage{footnote}
\usepackage[multiple]{footmisc}
\usepackage{textcomp}
\usepackage{xspace} 
\usepackage{booktabs}
\usepackage{multirow}
\usepackage[nointegrals]{wasysym}
\usepackage{scalerel}
\usepackage{xcolor}
\usepackage{mfirstuc}
\usepackage{glossaries-extra}
\usepackage{pdflscape}

\DeclareSIUnit\kB{\kilo\byte}
\DeclareSIUnit\MB{\mega\byte}
\DeclareSIUnit\GB{\giga\byte}
\DeclareSIUnit\TB{\tera\byte}
\DeclareSIUnit\kiB{\kibi\byte}
\DeclareSIUnit\MiB{\mebi\byte}
\DeclareSIUnit\GiB{\gibi\byte}
\DeclareSIUnit\TiB{\tebi\byte}
\DeclareSIUnit\kbit{\kilo\bit}
\DeclareSIUnit\Mbit{\mega\bit}
\DeclareSIUnit\Gbit{\giga\bit}
\DeclareSIUnit\Tbit{\tera\bit}
\DeclareSIUnit\kibit{\kibi\bit}
\DeclareSIUnit\Mibit{\mebi\bit}
\DeclareSIUnit\Gibit{\gibi\bit}
\DeclareSIUnit\Tibit{\tebi\bit}

\DeclareSIUnit[per-mode=symbol]\bps{\bit\per\second}
\DeclareSIUnit[per-mode=symbol]\kbps{\kilo\bit\per\second}
\DeclareSIUnit[per-mode=symbol]\Mbps{\mega\bit\per\second}
\DeclareSIUnit[per-mode=symbol]\Gbps{\giga\bit\per\second}
\DeclareSIUnit[per-mode=symbol]\Tbps{\tera\bit\per\second}
\DeclareSIUnit[per-mode=symbol]\Bps{\byte\per\second}
\DeclareSIUnit[per-mode=symbol]\kBps{\kilo\byte\per\second}
\DeclareSIUnit[per-mode=symbol]\MBps{\mega\byte\per\second}
\DeclareSIUnit[per-mode=symbol]\GBps{\giga\byte\per\second}
\DeclareSIUnit[per-mode=symbol]\TBps{\tera\byte\per\second}

\DeclareSIUnit[per-mode=symbol]\kiBps{\kiB\per\second}
\DeclareSIUnit[per-mode=symbol]\MiBps{\MiB\per\second}
\DeclareSIUnit[per-mode=symbol]\GiBps{\GiB\per\second}
\DeclareSIUnit[per-mode=symbol]\TiBps{\TiB\per\second}

\DeclareSIUnit[number-unit-product=\text{-}]\nbit{bit}
\DeclareSIUnit[number-unit-product=\text{-}]\nbyte{byte}
\newcommand*{\nbit}[1]{\SI{#1}{\nbit}}
\newcommand*{\nbyte}[1]{\SI{#1}{\nbyte}}

\DeclareSIUnit\clight{\text{\ensuremath{c}}}
\DeclareSIUnit{\sample}{S}

\newcommand{\murm}{%
  \ifmmode
    \mathchoice
        {\hbox{\normalsifze\textmu}}
        {\hbox{\normalsize\textmu}}
        {\hbox{\scriptsize\textmu}}
        {\hbox{\tiny\textmu}}%
  \else
    \textmu
  \fi
}

\newcommand{\pp}{\ensuremath{\mathrm {p\kern-0.05em p}}\xspace}
\newcommand{\PbPb}{\ensuremath{\mbox{Pb--Pb}}\xspace}

\newcommand{\figref}[1]{Fig.~\ref{#1}}
\newcommand{\Figref}[1]{Figure~\ref{#1}}	
\newcommand{\secref}[1]{Section~\ref{#1}}

\newcommand{\Tabref}[1]{Table~\ref{#1}}

\newcommand*{\eg}{e.g.,\xspace}
\newcommand*{\ie}{i.e.,\xspace}
\newcommand*{\cf}{cf.\xspace}
\newcommand*{\ebtb}{8b10b\xspace}

\newcommand*{\stage}[1]{Stage\,{#1}\xspace}

\newcommand*{\variable}[1]{{\small {\tt {#1}}}\xspace}

\setabbreviationstyle[acronym]{long-short}
\glssetcategoryattribute{acronym}{nohyperfirst}{true}

\newacronym{ASIC}{ASIC}{application-specific integrated circuit}
\newacronym{ADC}{ADC}{analog-to-digital converter}
\newacronym{ALICE}{ALICE}{A Large Ion Collider Experiment}
\newacronym{EPN}{EPN}{Event-Processing Node}
\newacronym{CDC}{CDC}{clock domain crossing}
\newacronym{CM}{CM}{common-mode value}
\newacronym{CRC}{CRC}{cyclic redundancy check}
\newacronym{CRU}{CRU}{Common Readout Unit}
\newacronym{CTP}{CTP}{Central Trigger Processor}
\newacronym{DAS}{DAS}{direct ADC sampling}
\newacronym{DCS}{DCS}{Detector Control System}
\newacronym{DSP}{DSP}{digital signal processor}
\newacronym{ELink}{E-link}{serial electrical link}
\newacronym{FPGA}{FPGA}{field-programmable gate array}
\newacronym{FEC}{FEC}{front-end card}
\newacronym{FEE}{FEE}{front-end electronics}
\newacronym{FLP}{FLP}{First-Level Processor}
\newacronym{GBT}{GBT}{Giga-Bit Transceiver}
\newacronym{GFD}{GFD}{\gls{GBT} frame decoder}
\newacronym{GBTx}{GBTx}{GBT transceiver \gls{ASIC}}
\newacronym{GEM}{GEM}{gas electron multiplier}
\newacronym{GPU}{GPU}{graphics processing unit}
\newacronym{HB}{HB}{heartbeat}
\newacronym{HBF}{HBF}{heartbeat frame}
\newacronym{HLT}{HLT}{High Level Trigger}
\newacronym{HVCM}{HVCM}{High Voltage Current Monitor}
\newacronym{IDC}{IDC}{integrated digital current}
\newacronym{IROC}{IROC}{Inner ReadOut Chamber}
\newacronym{ITF}{ITF}{ion tail filter}
\newacronym{LHC}{LHC}{Large Hadron Collider}
\newacronym{LTU}{LTU}{Local Trigger Unit}
\newacronym{OROC}{OROC}{Outer ReadOut Chamber}
\newacronym{PCIe}{PCIe}{PCI express}
\newacronym{PON}{PON}{passive optical network}
\newacronym{MWPC}{MWPC}{multi-wire proportional chamber}
\newacronym{RDH}{RDH}{Raw Data Header}
\newacronym{SerDes}{SerDes}{serializer/deserializer}
\newacronym{STF}{STF}{Sub-Time Frame}
\newacronym{SCA}{GBT-SCA}{GBT slow-control adapter}
\newacronym{SyncBox}{SyncBox}{Synchronisation Box}
\newacronym{TPC}{TPC}{Time Projection Chamber}
\newacronym{TPCUL}{TPC UL}{\gls{TPC} user logic}
\newacronym{TTC}{TTC}{Timing, Trigger and Control}
\newacronym{TTC-PON}{TTC-PON}{timing, trigger and control - passive optical network}
\newacronym{PLL}{PLL}{phase-locked loop}
\newacronym{TF}{TF}{Time Frame}
\newacronym{UL}{UL}{User Logic}
\newacronym{VTRx}{VTRx}{versatile transceiver}
\newacronym{VTTx}{VTTx}{versatile twin transmitter}
\newacronym{VLDB}{VLDB}{Versatile Link Demonstrator Board}
\newacronym{UART}{UART}{universal asynchronous receiver/transmitter}
\newacronym{SCN}{SCN}{slow control network}

\newacronym{TB}{time-bin}{time bin}\glsunset{TB}

\title{Large-scale real-time signal processing in physics experiments: The ALICE TPC FPGA pipeline}
\newcounter{daggerfootnote}

\newcommand*{\daggerfootnote}[1]{%
  \setcounter{daggerfootnote}{\value{footnote}}%
  \renewcommand*{\thefootnote}{\fnsymbol{footnote}}%
  \footnote[2]{#1}
  \setcounter{footnote}{\value{daggerfootnote}}%
  \renewcommand*{\thefootnote}{\arabic{footnote}}%
}

\newcommand{\daggermark}{%
  {\renewcommand{\thefootnote}{\fnsymbol{footnote}}\footnotemark[2]}%
}

\affiliation[a]{{Comenius University Bratislava, Faculty of Mathematics, Physics and Informatics}, {Bratislava}, {Slovakia}}
\affiliation[b]{{Department of Physics, University of Oslo}, {Norway}}
\affiliation[c]{{Department of Physics and Technology, University of Bergen}, {Norway}}
\affiliation[d]{{European Organization for Nuclear Research (CERN)}, {Geneva}, {Switzerland}}
\affiliation[e]{{Faculty of Engineering and Science, Western Norway University of Applied Sciences}, {Bergen}, {Norway}}
\affiliation[f]{{Helmholtz-Institut f\"{u}r Strahlen- und Kernphysik, Rheinische Friedrich-Wilhelms-Universit\"{a}t Bonn}, {Germany}}
\affiliation[g]{{Horia Hulubei National Institute of Physics and Nuclear Engineering}, {Bucharest}, {Romania}}
\affiliation[h]{{Institut f\"{u}r Kernphysik, Johann Wolfgang Goethe-Universit\"{a}t Frankfurt}, {Germany}}
\affiliation[i]{{Instituto de Ciencias Nucleares, Universidad Nacional Aut\'{o}noma de M\'{e}xico}, {Mexico City}, {Mexico}}
\affiliation[j]{{Lund University Department of Physics, Division of Particle Physics}, {Lund}, {Sweden}}
\affiliation[k]{{Nagasaki Institute of Applied Science}, {Nagasaki}, {Japan}}
\affiliation[l]{{Niels Bohr Institute, University of Copenhagen}, {Denmark}}
\affiliation[m]{{Oak Ridge National Laboratory}, {Oak Ridge}, {Tennessee}, {USA}}
\affiliation[n]{{Physics department, Faculty of science, University of Zagreb}, {Croatia}}
\affiliation[o]{{Physikalisches Institut, Ruprecht-Karls-Universit\"{a}t Heidelberg}, {Germany}}
\affiliation[p]{{Physik Department, Technische Universit\"{a}t M\"{u}nchen}, {Munich}, {Germany}}
\affiliation[q]{{Research Division and ExtreMe Matter Institute EMMI, GSI Helmholtzzentrum f\"ur Schwerionenforschung GmbH}, {Darmstadt}, {Germany}}
\affiliation[r]{{The Henryk Niewodniczanski Institute of Nuclear Physics, Polish Academy of Sciences}, {Cracow}, {Poland}}
\affiliation[s]{{The University of Texas at Austin}, {Texas}, {USA}}
\affiliation[t]{{Universidade de S\\,{a}o Paulo (USP)}, {Brazil}}
\affiliation[v]{{University of Houston}, {Texas}, {USA}}
\affiliation[w]{{University of Tennessee}, {Knoxville}, {Tennessee}, {USA}}
\affiliation[x]{{University of Tokyo}, {Japan}}
\affiliation[y]{{Wigner Research Centre for Physics}, {Budapest}, {Hungary}}
\affiliation[z]{{Yale University}, {New Haven}, {Connecticut}, {USA}}

\collaboration{ALICE TPC collaboration}
\author[c]{J.\,Alme}
\author[h,1]{T.\,Alt\note{Corresponding author.}}
\emailAdd{Torsten.Alt@cern.ch}
\author[g]{C.\,Andrei}
\author[o]{V.\,Anguelov}
\author[h]{H.\,Appelsh\"{a}user}
\author[z]{M.\,Arslandok}
\author[q]{R.\,Averbeck}
\author[f]{M.\,Ball}
\author[y]{G.\,G.\,Barnaf\"{o}ldi}
\author[o,q]{P.\,Becht}
% ...
\abstract{
  For LHC Run\,3, the ALICE Time Projection Chamber was upgraded to operate in continuous readout mode.
  Interaction rates of up to \SI{50}{\kilo\hertz} in \PbPb collisions require real-time processing of more than \SI{3}{\tera\byte\per\second} of raw detector data.
  This requirement is met by a custom FPGA-based processing pipeline that performs the complete front-end data treatment fully in-stream, including common-mode correction, pedestal subtraction, ion-tail filtering, zero suppression, and dense data packing.

  A central element of the design is a highly parallel common-mode correction algorithm operating directly on the streaming data.
  It robustly identifies signal-free readout channels on a time-bin basis and applies pad-dependent scaling to compensate for local variations in capacitive coupling in the GEM readout.
  In combination with pedestal subtraction and ion-tail filtering, this enables accurate baseline restoration under extreme high-occupancy conditions, preventing signal loss while efficiently suppressing noise prior to zero suppression.

  The pipeline operates continuously at the full detector bandwidth and reduces the raw input rate of approximately \SI{3}{\tera\byte\per\second} to about \SI{900}{\giga\byte\per\second} for \PbPb collisions at the target interaction rate.
  Overall, it represents a large-scale FPGA-based real-time signal-processing implementation for high-energy physics detector readout.
}
\arxivnumber{2601.15868}


\begin{document}
\maketitle
\flushbottom

\section{The upgraded ALICE TPC}
\label{sec:TPC}

The \gls{TPC} of \gls{ALICE}, located at the CERN \gls{LHC}, served as the main tracking detector during Runs~1 and~2 (2009–2018), employing \glspl{MWPC} with cathode-pad signal readout~\cite{TPCnim}.

To meet the demands of Runs\,3 and 4, the \gls{TPC} underwent a major upgrade during the \gls{LHC} long shutdown from 2019 to 2021.
The upgrade involved replacing the \glspl{MWPC} with new readout chambers based on \gls{GEM} technology and installing entirely new \gls{FEE} in order to continuously read the signals on the anode pads~\cite{TPCjinst}.
This modernization was driven by the need to accommodate much higher interaction rates, reaching up to \SI{50}{\kilo\hertz} in lead–lead (\PbPb) collisions.

The \gls{ALICE} \gls{TPC} is a large cylindrical detector, divided longitudinally into two halves by a central drift electrode.
Each end plate, located at the ends of the cylinder, houses the readout chambers and is subdivided into 18 azimuthal sectors.
Every sector contains \glspl{IROC} and \glspl{OROC}.
The new GEM-based readout chambers use a stack of four \gls{GEM} foils for gas amplification.
Unlike the previous design, the upgraded system does not use an ion gating grid.
As a result, ion backflow---\ie the flow of positive ions into the drift volume---must be minimized.
This is achieved through a specially optimized amplification structure and a thoroughly tuned high-voltage configuration across the \gls{GEM} stack (\cf~\cite{TPCjinst}).

The upgraded \gls{TPC} has been successfully commissioned and is now fully operational, effectively supporting the physics objectives of \gls{ALICE} in high-luminosity heavy-ion collisions~\cite{ALICELOI,zitrone}.
In this publication, we describe in detail the readout system of the \gls{ALICE} \gls{TPC}, with a particular focus on the data processing algorithms implemented in firmware on the readout cards that receive data from the \gls{TPC} \gls{FEE}.

\subsection{Overview on the data processing chain}
\label{sec:TPC:overview}

\begin{figure}[ht]
    \centering
    \includegraphics[width=0.85\linewidth]{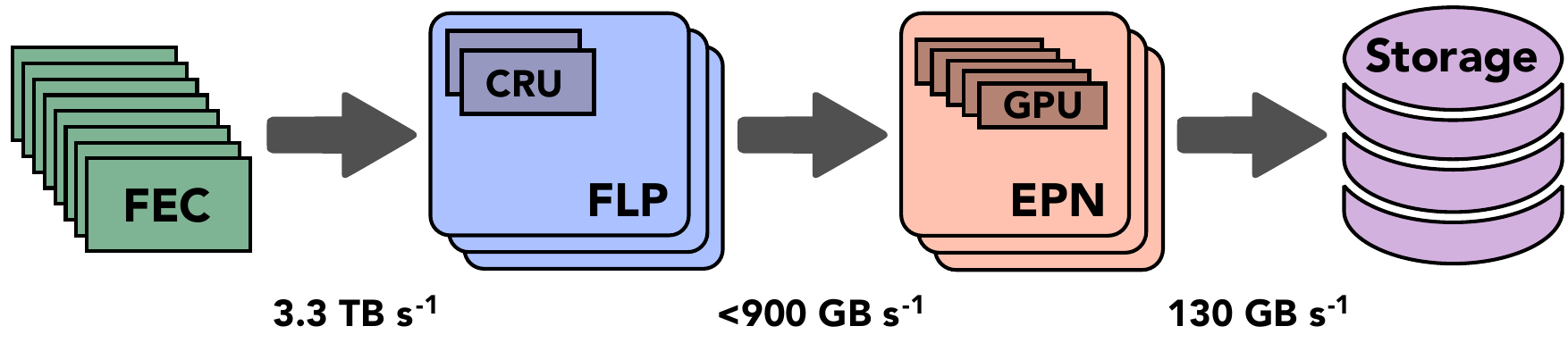}
    \caption{\gls{TPC} data processing chain.
    The front-end electronics continuously transmit all sampled data off-detector.
    In the First-Level Processors (FLP), each hosting two or three FPGA-based readout cards, known as Common Readout Units (CRUs), the data are corrected for specific detector effects and reduced in size.
    The resulting reduced data stream is distributed to the Event-Processing Nodes (EPNs).
    Each EPN performs online calibration, reconstruction, data reduction and data compression, before the data are directed to permanent storage.}
    \label{fig:TPC:scheme}
\end{figure}

The full data processing chain is illustrated schematically in \figref{fig:TPC:scheme}.
At the \gls{FEE} (\cf \secref{sec:TPC:FEC}), signals from the \num{524160} anode readout pads are amplified, shaped, and continuously sampled and digitized at a rate of \SI{5}{\mega\hertz}.
The digitized data are multiplexed into \num{6552} radiation-tolerant optical readout links.
At this stage, no data reduction is applied, ensuring that all raw detector information is available for subsequent correction of detector-specific effects at the next stage.
The total data rate from the \gls{TPC} amounts to \SI{3.3}{\tera\byte\per\second}.
The data are received by a total of 360 \glspl{CRU} (\cf \secref{sec:TPC:CRU}), each  containing a large \gls{FPGA}, which is programmed to perform data processing and reduction tasks (\cf \secref{sec:UL}).

The performance of the \gls{GEM}-based signal amplification is affected by two effects:
(i) baseline fluctuations induced by the \emph{\textit{common-mode effect}}, and
(ii) characteristic distortions in the pulse shape caused by slowly drifting ions in the induction gap (\emph{\textit{ion tails}}).
Both phenomena are described in \cite{CMCITpaper} and are corrected in the \gls{CRU} \gls{FPGA}.
Zero-suppression is applied only after these corrections, so that the data volume is effectively reduced without loss of information.

The \glspl{CRU} are installed in 144 \glspl{FLP}, which are dedicated computers, each hosting two or three \glspl{CRU}, depending on data throughput.
Data are pushed from the \glspl{FLP} to the \glspl{EPN} farm via custom distribution software deployed on both farms.
For \PbPb collisions at \SI{50}{\kHz}, the total data rate remains below \SI{900}{\giga\byte\per\second}.

The continuous stream is segmented into \glspl{TF} of configurable length (currently \SI{2.85}{\ms}).%
\footnote{The time frames are sized to fit in the \gls{GPU} memory for efficient processing on the \glspl{EPN}.}
Each \gls{FLP} produces \glspl{STF} from a TPC subset.
While not all \glspl{EPN} receive all \glspl{STF}, each one receives the complete set required to assemble a full \gls{TF}.

Since \gls{TPC} calibration and compression require track information, \glspl{EPN} perform full online track reconstruction during data taking.
Each server is equipped with eight \glspl{GPU}, which handle tracking and calibration entirely.

The output stream to storage is about \SI{130}{\giga\byte\per\second} for \PbPb at \SI{50}{\kHz}.
Further asynchronous processing, on the \glspl{EPN} or on the \gls{ALICE} computing grid,
refines the data to final precision.

\subsection{Front-end electronics}
\label{sec:TPC:FEC}

\begin{figure}
    \centering
    \includegraphics[width=0.7\linewidth]{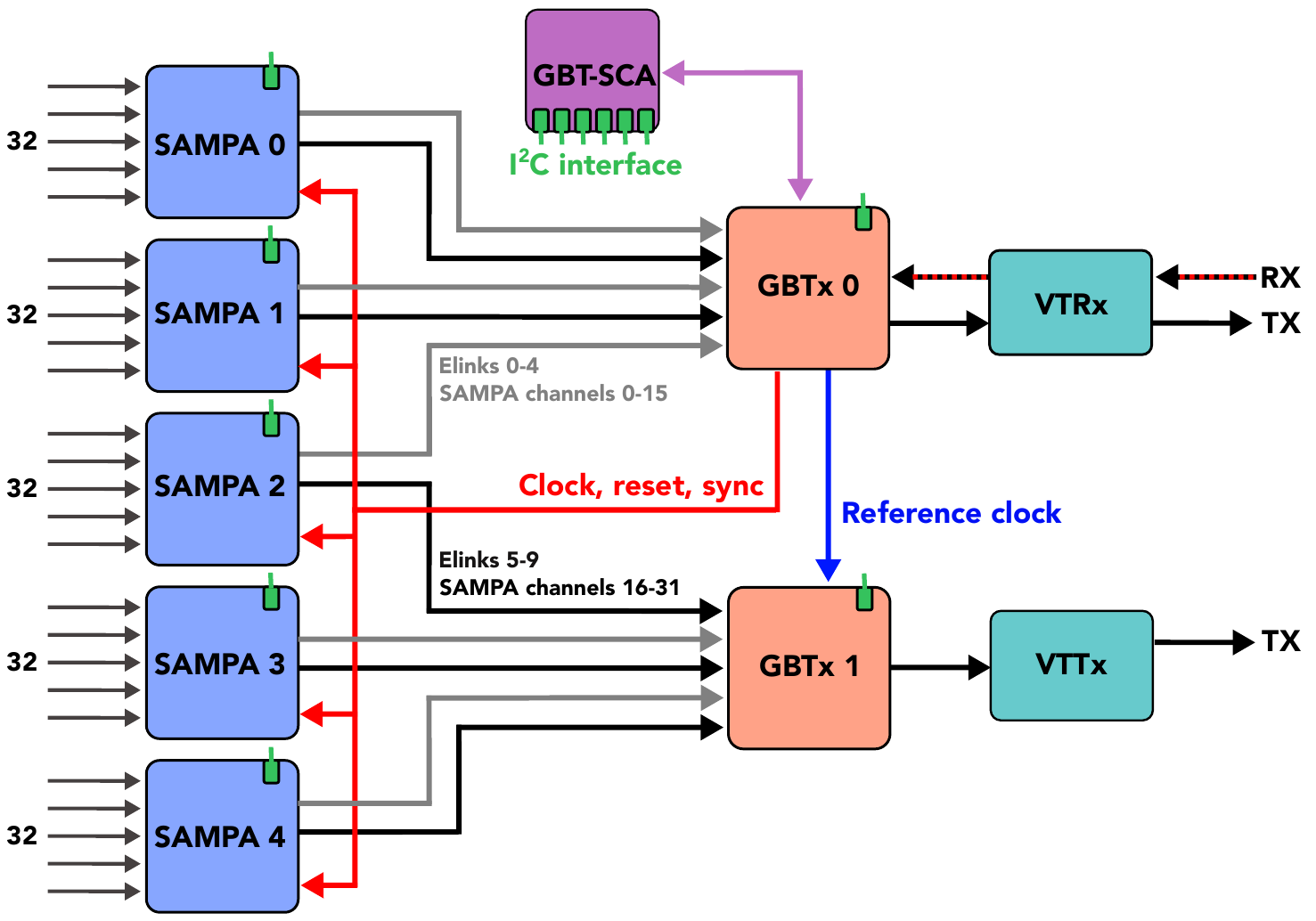}
    \caption{Schematic of the \gls{TPC} Front-End Card (FEC).
    Five \num{32}-channel SAMPA chips process pad signals and send them to two \gls{GBTx} chips for multiplexing and off-detector transfer.
    The SAMPA chips are operated in split mode, where the data from two groups of channels can be sent to different destinations.
    The \gls{FEC} connects via three optical fibers, two Tx for data and one Rx for commands, configuration, and clock.
    The clock is distributed from \gls{GBTx}0.
    All configuration is handled by the \gls{SCA} over I$^2$C.}
    \label{fig:TPC:fec}
\end{figure}

The \gls{FEC} for the \gls{ALICE} \gls{TPC} is described in detail in~\cite{TPCjinst} and illustrated schematically in \figref{fig:TPC:fec}.
It comprises the front-end \gls{ASIC} SAMPA, the \gls{GBT} chipset and Versatile Link components, along with auxiliary circuitry such as voltage regulators and temperature sensors.

\subsubsection{SAMPA}
\label{sec:TPC:FEC:SAMPA}

The SAMPA~\cite{sampa} \gls{ASIC} was specifically designed for the upgraded \gls{FEE} of both the \gls{TPC} and the \gls{ALICE} Muon Chambers~\cite{AliceMuon}.
The \gls{GBT} and Versatile Link components were developed at CERN to provide a radiation-tolerant, bidirectional optical link for data transmission as well as timing and control between the detector and off-detector systems.

The SAMPA is implemented in \SI{130}{\nm} CMOS technology with a nominal supply voltage of \SI{1.25}{\V}.
It features 32 channels with configurable input polarity and offers three combinations of shaping time and gain.
Each channel includes a charge-sensitive amplifier, a semi-Gaussian shaper, and a \nbit{10} \gls{ADC}.
A \gls{DSP} is integrated to provide digital filtering and data compression.
Data output is handled via 11 \glspl{ELink}, each operating at a configurable frequency.

Each \gls{FEC} hosts five SAMPA chips, providing a total of 160 readout channels.
In operation, the SAMPA is configured in \gls{DAS} mode, wherein the \gls{DSP} is bypassed and powered down.
In this mode, all \gls{ADC} data are streamed continuously, and the output contains no data markers.
Each SAMPA generates a data rate of \SI{1.6}{\giga\bit\per\second},
transmitted via ten \glspl{ELink} at \SI{160}{\mega\bit\per\second} each.%
\footnote{$\num{32}\,\textnormal{channels} \times \SI{10}{\bit} \times \SI{5}{\MHz} = \SI{1.6}{\giga\bit\per\second}$}
The 11th \gls{ELink} carries the \SI{5}{\MHz} \gls{ADC} sampling clock, allowing verification of alignment and phase consistency across the full \gls{TPC}.
This results in a total per-chip data rate of \SI{1.76}{\Gbps}.%
\footnote{$11\,\textnormal{\glspl{ELink}} \times \SI{160}{\mega\bit\per\second} = \SI{1.76}{\giga\bit\per\second}$}

The SAMPA operates in split mode, whereby the 32 readout channels are divided into two groups (0–15 and 16–31), each of which may be sent to a dedicated upstream receiver.
\glspl{ELink} 0–4 carry data from the first group, and \glspl{ELink} 5–9 from the second. 
Each 16-channel group is serialized over 32 \gls{ELink} clock cycles.
In each pair of cycles, the lower and upper 5 bits of an \gls{ADC} value are transmitted in parallel.
For example, during the first cycle, the lower 5 bits of channel 0 and channel 16 are transmitted simultaneously; the next cycle transmits the upper 5 bits, and the process continues channel by channel.

\subsubsection{Giga-Bit Transceiver}
\label{sec:TPC:FEC:GBT}

Two \gls{GBTx} chips, \gls{GBTx}0 and \gls{GBTx}1, multiplex the data and forward them to a \gls{VTRx} and a \gls{VTTx}.
The \gls{VTRx} is a bidirectional optical transceiver, while the \gls{VTTx} contains two transmit channels, one of which is used.
Thus, each \gls{FEC} employs two optical Tx fibers for data transmission and one Rx fiber for receiving clock, configuration, and control signals.

Each \gls{GBTx} receives data from two full SAMPAs.
The fifth SAMPA routes the first channel group to \gls{GBTx}0 and the second to \gls{GBTx}1.
The 11th \gls{ELink} from this chip is split to deliver the sampling clock to both \gls{GBTx} devices.

The total input data rate per \gls{GBTx} is \SI{4.48}{\giga\bit\per\second}.%
\footnote{$(\num{5}\times\num{5}+\num{3})\,\textnormal{\glspl{ELink}} \times \SI{160}{\mega\bit\per\second} = \SI{4.48}{\giga\bit\per\second}$}
The Versatile Link operates at a line rate of \SI{4.8}{\giga\bit\per\second} and delivers in wide-frame mode%
\footnote{The standard \gls{GBT} frame format includes forward error correction, which reduces the effective data rate to \SI{3.2}{\giga\bit\per\second}.
To achieve higher throughput, the wide-frame mode disables forward error correction, thereby increasing the usable data rate at the expense of error protection.
In the case of \gls{ALICE}, this trade-off is acceptable: the radiation background is significantly lower than in typical Versatile Link applications, so the absence of forward error correction does not pose a concern for data integrity.}
as usable data bandwidth exactly the needed \SI{4.48}{\giga\bit\per\second}.
The total data rate per \gls{FEC} is thus \SI{8.96}{\giga\bit\per\second}, excluding monitoring overhead.

The Rx link from the \gls{CRU} to the \gls{FEC} has fixed and deterministic latency and is protected by forward error correction.
It is used to transmit clock, control, and configuration data via the \gls{VTRx} to \gls{GBTx}0.

Finally, among the key components on the \gls{FEC} is the \gls{SCA} chip, responsible for configuration and monitoring.
It communicates with other components on the board using the I\textsuperscript{2}C protocol.

\subsection{Common readout unit}
\label{sec:TPC:CRU}

The common readout unit (CRU) is a custom \gls{FPGA} card~\cite{pcie40} developed collaboratively by LHCb~\cite{LHCbupgrade} and \gls{ALICE}~\cite{TDR:rdo}.
It follows a \gls{PCIe} form factor (three-quarter length, standard height, dual-slot) and complies with \gls{PCIe} specifications.
The board is based on an Intel\textsuperscript{TM} Arria 10 \gls{FPGA} featuring \num{72} \gls{SerDes}, of which \num{16} are dedicated to the host interface with the PC.
The \gls{CRU} also provides an interface to the \gls{TTC-PON} for receiving trigger and timing information distributed by the experiment.
The \gls{FEE} are connected via up to \num{48} bi-directional optical links, each supporting bandwidths of up to \SI{10}{\giga\bit\per\second} in both directions.

The \gls{CRU} firmware consists of a \emph{\textit{Common Logic}}%
\footnote{Maintained by a central development team.}
including common design components shared across several \gls{ALICE} subdetectors and a subdetector-specific \emph{\textit{\gls{UL}}}.
For the \gls{TPC}, two distinct top-level design \gls{UL} units have been implemented, each serving fundamentally different purposes:

\begin{itemize}
\item \num{360} \glspl{CRU} operate with the data-processing \gls{UL} to handle collision data and transmit it to the \glspl{FLP} (\cf \secref{sec:UL}), and
\item an additional \gls{CRU} performs synchronized control of the \gls{TPC} laser and pulser systems, and reads out the monitoring system of the \gls{GEM} currents (\cf \secref{sec:ULSync}).
\end{itemize}

Since the latter functionality requires only modest logic resources, both design units are integrated into a single \gls{TPCUL}, simplifying firmware build and deployment.
The operational mode---\ie which of the two design units is active---can be selected as needed.
In the remainder of this paper, these two configurations are treated as distinct \gls{UL} flavors.

\subsection{Timing}
\label{sec:TPC:Timing}

At the \gls{LHC}, the fundamental units of time are defined by the orbit number and the bunch crossing number.
An orbit corresponds to the time it takes for a particle bunch to complete a full revolution around the accelerator ring: approximately \SI{89.1}{\micro\second}.
A bunch crossing refers to the event in which two particle bunches, traveling in opposite directions, pass through an interaction point.%
\footnote{
In actual beam operation, not all bunches are filled with particles, and therefore not every bunch crossing results in a collision.}
The bunch crossing rate, approximately \SI{40}{\mega\hertz}, defines the fundamental clock frequency distributed from the \gls{LHC} to the experiments.
All accelerator operations, data acquisition systems, and detector electronics are synchronized based on these time units.

\begin{figure}[ht]
    \centering
    \includegraphics[width=0.80\linewidth]{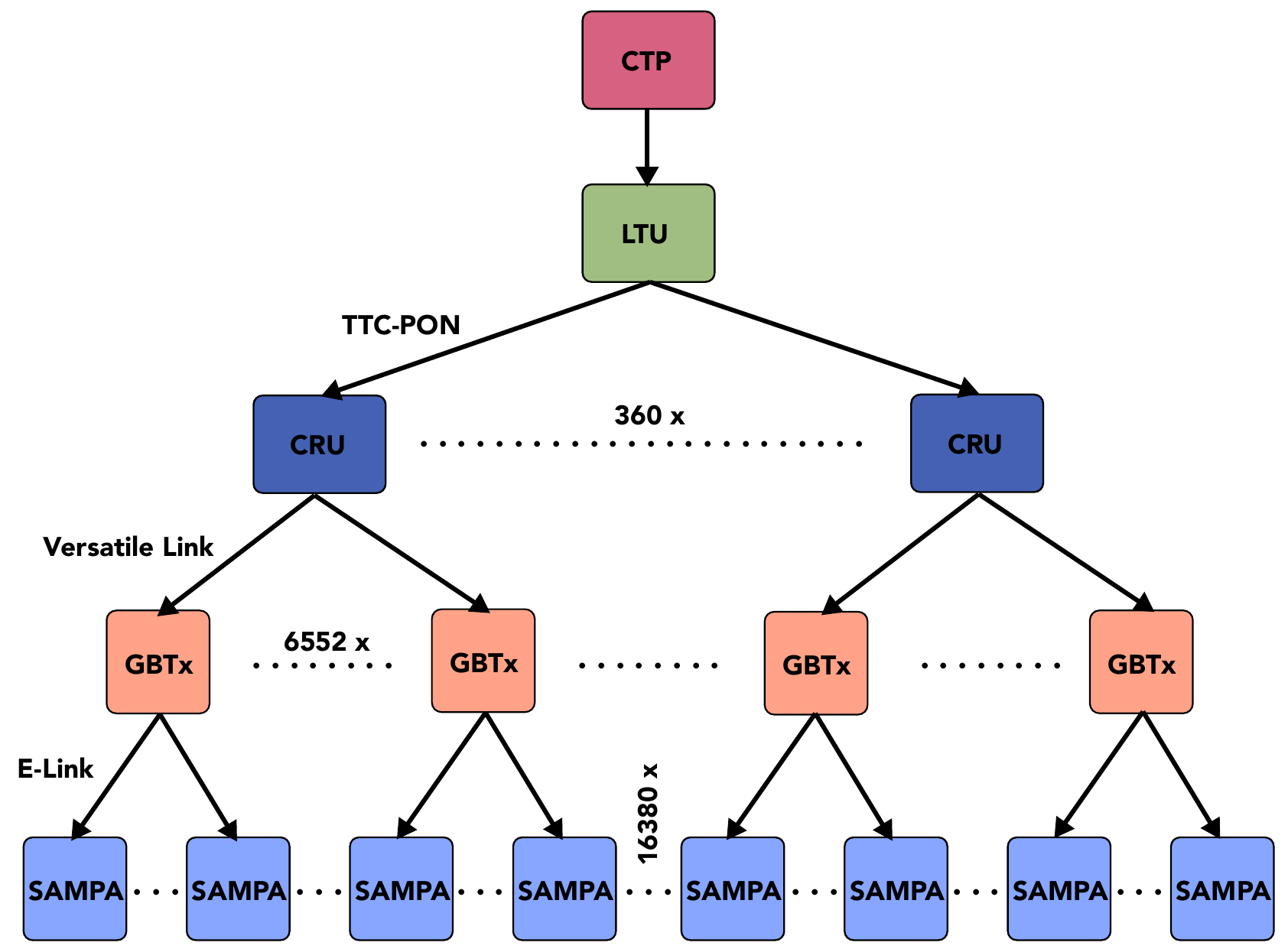}
    \caption{ Clock and trigger distribution tree for the \gls{ALICE} \gls{TPC}. }
    \label{fig:TPC:Clock_Trigger}
\end{figure}

In \gls{ALICE}, the distribution of the bunch crossing clock, along with crucial timing information such as orbit and bunch crossing counters%
\footnote{After each completed orbit, corresponding to \num{3564} bunch crossings, the bunch-crossing counter resets to \num{0}, and the orbit counter increments by one.}
and trigger signals, is managed by the \gls{TTC} system.
\Figref{fig:TPC:Clock_Trigger} illustrates the clock and trigger distribution network for the \gls{TPC}, which forms part of the global ALICE \gls{TTC} infrastructure.

The \gls{CTP} serves as the central synchronization node for \gls{ALICE}, interfacing with the \gls{LHC} to generate the clock, timing, and trigger signals used across the experiment.
This information is passed to the detector-specific \gls{LTU}, which encodes it into a high-speed serial signal.
The signal is then distributed to all \glspl{CRU} via the \gls{TTC-PON}.

Each \gls{CRU} recovers the distributed clock and extracts the trigger message.
The recovered clock is phase-synchronous across all \glspl{CRU} and is used to drive communication between the \glspl{CRU} and the \gls{GBTx}0 chips on the \glspl{FEC}.
Each \gls{GBTx}0 generates the \SI{160}{\mega\hertz} \gls{ELink} clock, which is then distributed to the SAMPA chips.
The SAMPA chips internally derive the \SI{5}{\mega\hertz} \gls{ADC} sampling clock from the \gls{ELink} clock using a clock divider.%
\footnote{This approach avoids the need for a phase-locked loop (PLL), which is less reliable in radiation-prone environments.}

A key characteristic of the Versatile Link and the \gls{GBTx} is their guaranteed fixed and deterministic latency on the Rx path to the \gls{FEC}.
This ensures that signals transmitted over the Versatile Link arrive with identical timing on the \gls{ELink} pins of all \gls{GBTx} devices across the \glspl{FEC}, thereby enabling system-wide synchronization.

The trigger message sent from the \gls{LTU} and received by the \gls{CRU} at the \SI{40}{\mega\hertz} bunch crossing rate includes orbit and bunch-crossing numbers along with trigger bits.
This information is not forwarded to the \glspl{FEC}.
Instead, the timing-critical operation of the SAMPAs is controlled directly by the \gls{CRU} via the fixed-latency path described above, using dedicated signal lines between the corresponding SAMPA and \gls{GBTx} pins (\cf \secref{sec:TPC:Synchronisation}).

The data collected from the input links are segmented into time slices called \glspl{HBF}.
An \gls{HBF} corresponds to the data bounded by two heartbeat signals (HB), which are periodic signals synchronized to the LHC orbit at a frequency of \SI{11.223}{\kilo\hertz}.
An \gls{HBF} represents the smallest unit of data processed at the \gls{FLP} level.
Each \gls{HBF} is tagged with an HB ID, which includes the orbit number, bunch crossing number, and trigger information.

At the \gls{FLP}, a configurable number of consecutive \glspl{HBF} (currently 32) are merged to form an \gls{STF}.
At the \glspl{EPN}, all \glspl{STF} from the various \glspl{FLP} that correspond to the same time period are aggregated into a \gls{TF} (\cf \secref{sec:TPC:overview}).

\subsection{Synchronization of ADC sampling clocks and data streams}
\label{sec:TPC:Synchronisation}

The SAMPA chips generate the \gls{ADC} clock internally from the phase-synchronous \SI{160}{\mega\hertz} \gls{ELink} clock using a simple clock divider within each SAMPA chip.
Since this divider begins counting immediately upon power-up, the resulting \gls{ADC} clock phase can differ between SAMPAs, even between those on the same FEC.

To synchronize  the \gls{ADC} clocks across all SAMPA chips on the \gls{TPC}, each SAMPA provides a physical \textit{RESET} pin, which is directly connected to an \gls{ELink} on the \gls{GBTx}0.
This \gls{ELink} is controlled by the \gls{CRU}, and due to the fixed-latency characteristics of the \gls{GBT} system, the timing of the \textit{RESET} signal is deterministic.
This allows the \gls{CRU} to synchronously reset all SAMPAs connected to it.

For global synchronization of the entire \gls{TPC}, a reference signal from the \gls{TTC} system can be employed.
Suitable \gls{TTC} references include a dedicated \gls{TPC} \textit{RESET} trigger, a specific bunch-crossing identifier, or every $N$th \gls{LHC} orbit.
Any such reference can be mapped inside the \gls{CRU} to generate a synchronized \textit{RESET}, which is then distributed to all connected SAMPA chips.
A dedicated module in the Common Logic, the Pattern Player, performs this mapping by converting the selected \gls{TTC} reference into predefined pattern sequences that are subsequently transmitted to the SAMPAs.

As detailed in \secref{sec:TPC:FEC}, the SAMPA chips operate in \gls{DAS} mode, in which the stream of \gls{ADC} samples contains no embedded metadata or framing markers.
Instead, the mapping between each \gls{ADC} value and its corresponding channel is inferred solely from its position in the continuous output stream.
To ensure correct alignment, a dedicated \textit{SYNC} pattern is transmitted at the beginning of each stream, marking its start.
After sending this pattern, the SAMPA enters a deterministic mode in which it cyclically transmits the \gls{ADC} values from all input channels in a fixed order.
One full cycle of these channel-ordered \gls{ADC} values is referred to as a \emph{\textit{time-bin}}.

Upon issuing a \textit{RESET}, the SAMPA automatically sends the \textit{SYNC} pattern at the beginning of the next data transmission.
Additionally, the \textit{SYNC} pattern can be triggered manually via a dedicated \textit{ENABLE} pin on the SAMPA, which, like the \textit{RESET} pin, is connected to an \gls{ELink} and thus controllable via the \gls{CRU}.

For details on re-synchronization during decoding of the received data in the \gls{TPCUL}, refer to \secref{sec:UL:Decode:Resync}.

\subsection{Data format of the GBT frames}
\label{sec:TPC:GBTframes}

Each \gls{FEC} uses two \gls{GBTx} chips to collect, arrange, and transmit the data from its five connected SAMPA chips.
The outgoing data are packaged into \gls{GBT} frames, each consisting of 120 bits.
In wide-frame mode (\cf \secref{sec:TPC:FEC:GBT}), 112 of these bits are available for user data.%
\footnote{$112\,\textnormal{bits} \times \SI{40}{\mega\hertz}\ \textnormal{(LHC clock frequency)} = \SI{4.48}{\giga\bit\per\second}$}
The \gls{GBTx} chips acquire data from 28 connected \gls{ELink} inputs operating at \SI{160}{\mega\hertz}.
Each \gls{GBT} frame thus encapsulates four consecutive bits per \gls{ELink}.%
\footnote{$(5 \times 5 + 3)\,\text{E-links} \times 4\,\text{bits} = 112\,\textnormal{bits}$}

To encode the 32 \gls{ADC} samples generated in a single \gls{ADC} clock cycle (one time-bin), eight \gls{GBT} frames are required.%
\footnote{$8 \times 112\,\textnormal{bits} \times \SI{5}{\mega\hertz}$ (\gls{ADC} sampling clock) = \SI{4.48}{\giga\bit\per\second}}
Within the \gls{CRU} firmware, these \gls{GBT} frames are demultiplexed and unpacked on a per-link basis (\cf \secref{sec:UL:Decode:Decode}).

\subsection{Readout modes}
\label{sec:TPC:Readoutmodes}

Although the data stream from \gls{FEE} to the \glspl{CRU} is inherently continuous, the \gls{CRU} supports two distinct modes for forwarding data to the \glspl{FLP}.
In \textit{continuous readout mode}, all data are transmitted to the \gls{FLP}, with triggers embedded as metadata for reference during downstream processing.
In \textit{triggered readout mode}, data are transmitted to the \gls{FLP} only for predefined time windows, and only in response to valid triggers.
Physics data taking with collisions is done in continuous readout mode, while calibration runs, e.g. to deterime pedestal values and noise, are done in triggered readout mode.
\section{User logic for data processing}
\label{sec:UL}

The \gls{TPCUL} forms the subdetector-specific part of the 
\gls{CRU} \gls{FPGA} firmware and is responsible for processing raw 
\gls{TPC} data.
Within the \gls{UL}, the fundamental data unit is a single 
\gls{ADC} value acquired from one SAMPA channel in a given time-bin.
In the following, such an \gls{ADC} value is referred to as a \emph{\textit{sample}}.

During processing, the samples are sequentially propagated through multiple firmware stages, where correction operations are applied.
To improve numerical precision in these calculations, the \nbit{10} resolution of the SAMPA chip \gls{ADC} is extended to 12 bits by adding two least-significant bits in the firmware.

The \gls{UL} architecture is constrained by the data throughput requirements and available \gls{FPGA} resources.
It must sustain the processing of up to 1600 channels, each generating a new sample every \SI{200}{\nano\second} (\SI{5}{\mega\hertz} \gls{ADC} sampling rate).
In addition, it must implement multiple correction algorithms optimized for \gls{TPC} operational conditions.
These algorithms are classified according to processing scope:

\begin{enumerate}
    \item Per-channel, multi-time-bin operations – Each channel is processed independently across time bins, often requiring feedback to maintain state between iterations.
    This category includes digital filtering operations, e.g., the ion-tail filter (\cf \secref{sec:UL:Blocks:ITF}).
    \label{itemone}
    \item Multi-channel, single-time-bin operations – Multiple channels are processed concurrently at a fixed time bin, with no temporal dependency.
    An example is the common-mode filter (\cf \secref{sec:UL:Blocks:CMC}).
    \label{itemtwo}
    \item Single-channel, single-time-bin operations – Local transformations are applied to individual samples without inter-channel or temporal references.
    Examples include pedestal correction and the threshold filter (\cf \secref{sec:UL:Blocks:PedZS}).
    \label{itemthree}
\end{enumerate}

A central design challenge was the definition of a data format supporting efficient execution across all three categories.
While operations of type~(\ref{itemthree}) are straightforward, types~(\ref{itemone}) and~(\ref{itemtwo}) require conflicting data access patterns---temporal versus spatial---necessitating a flexible yet efficient data representation.

Given the suitability of \glspl{FPGA} for highly parallel, streaming-based processing pipelines, the adopted approach maintains continuous data flow through the processing chain, enabling real-time sample modification.
The following subsection describes the data format devised to meet these requirements.

\subsection{Data format of the processing pipeline}
\label{sec:UL:Dataformat}

\begin{figure}
  \centering
  \includegraphics[width=0.7\linewidth]{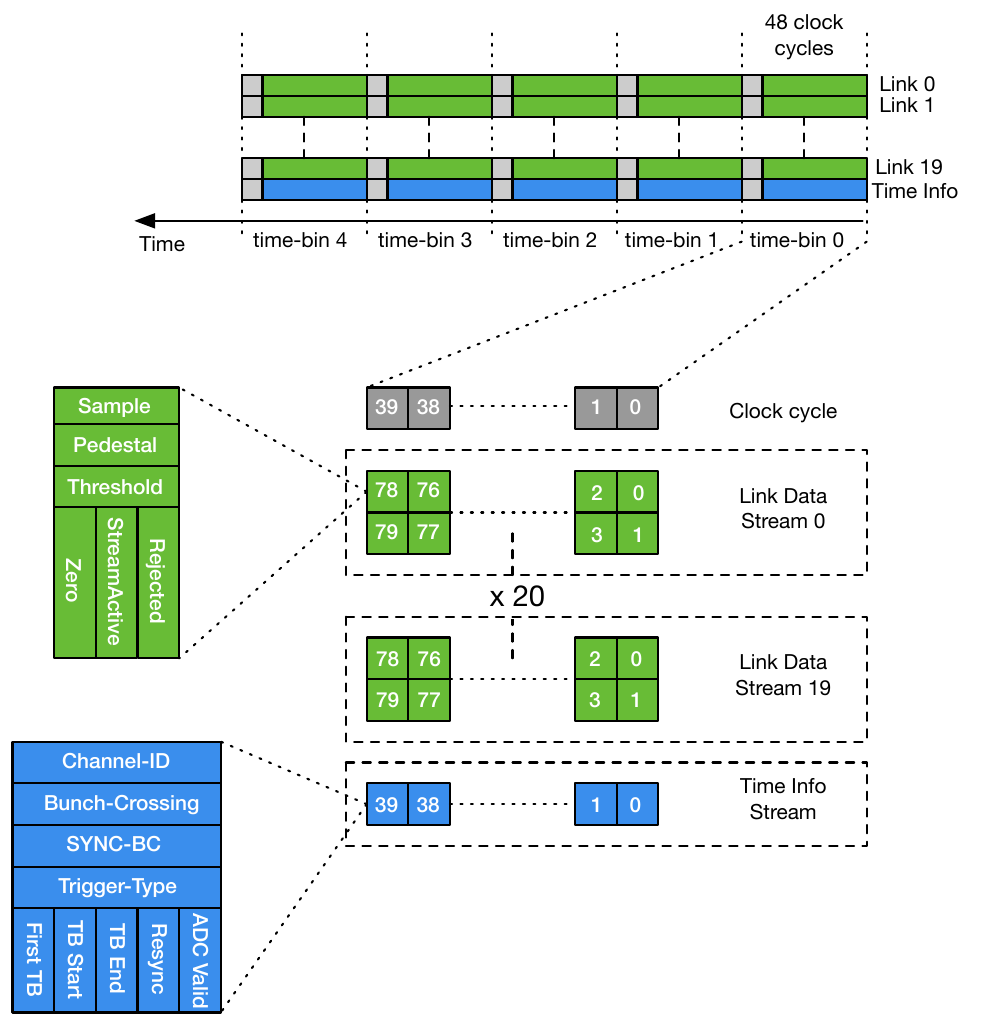}
  \caption{Data format of the processing pipeline. All flags and other information are described in the text. }
  \label{fig:UL:data_format_pipeline}
\end{figure}

The \gls{UL} operates at a nominal frequency of \SI{240}{\mega\hertz}, while up to \num{1600} samples are continuously acquired at \SI{5}{\mega\hertz}.
The \num{1600} new samples per time-bin must be accepted and processed continuously within the corresponding \num{48} clock cycles.
This requirement constrains the sustained processing throughput.
The latency between input and output, on the other hand, is not critical, provided that the pipeline can sustain the input rate without data loss.

Data are received over up to \num{20} optical links, each carrying a maximum of \num{80} channels.
To meet the throughput requirement, at least two samples per clock cycle per link must be processed.
With this scheme, all channel data are ingested within \num{40} cycles, leaving an \num{8}-cycle margin for auxiliary operations such as state management and synchronization.

For efficient, pipelined execution, the data format is designed to be compact and self-contained.
Each sample is accompanied by channel-specific parameters—including pedestal and threshold values—and a set of \nbit{1} flags conveying sample state information.
These flags enable localized decision-making without requiring access to external context. This dataset consisting of sample value, key parameters and flags is called \textit{\emph{link data stream}}.

Processing efficiency is further improved through temporal alignment of the \num{20} link data streams.
This alignment renders certain metadata, such as timing information and channel identifiers, common across all streams.
Such shared information is extracted and consolidated into a dedicated \textit{\emph{time info stream}}, which is propagated in parallel with the link data streams throughout processing.

The resulting format ensures that each clock cycle conveys all metadata and control signals necessary for sample processing.
Control and steering flags, as well as auxiliary parameters such as pedestal and threshold values, are embedded alongside the sample data to support real-time correction and filtering operations.

\Figref{fig:UL:data_format_pipeline} illustrates the architecture of this data format.
The pipeline comprises \num{20} link data streams and one time info stream.
Each link data stream transmits data for two channels per clock cycle. For each channel, the data includes these three flags:

\begin{itemize}
    \item \textit{Zero}: Indicates that the sample (after pedestal subtraction) is zero and may be suppressed,
    \item \textit{StreamActive}: Denotes that the data stream is currently active and synchronized;
    if a link or an \gls{FEC} is lost, this flag is cleared and the processing logic may react accordingly,
    \item \textit{Rejected}: Indicates that the sample should not be forwarded to the \gls{FLP};
    however, the sample remains available for internal processing (e.g., ion-tail filter).
\end{itemize}

The shared time info stream provides global metadata and steering information common to all link data streams, including these flags:

\begin{itemize}
    \item \textit{Channel-ID}: Identifies which channels are present in the current clock cycle,
    \item \textit{Bunch-Crossing}: Provides a timestamp corresponding to the \gls{LHC} bunch-crossing for the sample,
    \item \textit{Sync-BC}: Indicates the bunch-crossing number when the most recent \textit{SYNC} pattern was received,
    \item \textit{Trigger-Type}: Encodes any trigger information associated with the current time-bin,
    \item \textit{TB Start} and \textit{TB End}: One-bit flags indicating the start and end of a time-bin sample stream, respectively,%
    \footnote{The start and end could also be inferred from the \textit{Channel-ID}.
    However, providing explicit flags avoids the need for costly 6-bit comparator logic in each processing module.}
    \item \textit{ADC valid}: Indicates whether valid sample data is present in the current clock cycle (eight of the 48 cycles typically do not contain valid data),
    \item \textit{First TB}: Marks the first time-bin after the start of a run or a successful re-synchronization (\cf \secref{sec:UL:Decode:Resync});
    this flag can be used to initialize or reset processing components,
    \item \textit{Resync}: Signals that the system is currently undergoing re-synchronization (\cf \secref{sec:UL:Decode:Resync}), prompting the appropriate handling of incoming data.
\end{itemize}

This data format enables efficient processing within the \gls{UL}, supporting all three types of processing operations outlined in the previous section: per-channel temporal filters, per-time-bin spatial filters, and localized single-sample transformations.

\subsection{Resource utilisation}
\label{sec:UL:Resource}

\begin{table}[ht]\footnotesize
    \centering
    \caption{Resource utilisation of the Aria10 FPGA.}
    \begin{tabular}{l|ccc}
       \toprule
        Module & ALMs & M20K & DSP \\
        \midrule
        Available resources & 427.200 & 2.713 & 1.518\\
        Resources utilization & 303.601 & 1.424 & 242 \\
        Common Logic & 171.109 & 1.193 & 0\\
        User Logic & 132.492 & 231 & 242\\
        \midrule
        UL-Syncbox & 6.621 & 35 & 0\\
        GBT Frame Decoder & 15.905 & 0 & 0\\
        Global Aligner & 925 & 20 & 0\\
        Common-mode correction & 28.595 & 4 & 22\\
        Pedestal/Threshold memory & 599 & 40 & 0\\
        Pedestal subtraction and zero suppression & 6850 & 0 & 20\\
        Ion-Tail Filter & 18.501 & 0 & 200\\
        IDC processor & 5997 & 14 & 0\\
        Dense packing & 26.800 & 116 & 0\\
        Glue logic and monitoring & 21.699 & 2 & 0\\
        \bottomrule
    \end{tabular}
    \label{tab:fpga_resources}
\end{table}

\Tabref{tab:fpga_resources} summarizes the available FPGA resources and the utilization achieved by the implemented design on the Intel\textsuperscript{TM} Arria 10 device.
The reported metrics include Adaptive Logic Modules (ALMs), which implement combinational and sequential logic; M20K blocks, providing embedded on-chip memory; and DSP blocks, used for high-performance arithmetic operations such as multiplication and accumulation.

\subsection{Overview of the UL processing pipeline}
\label{sec:UL:ProcPipe}

\begin{figure}
  \centering
  \includegraphics[width=0.99\linewidth]{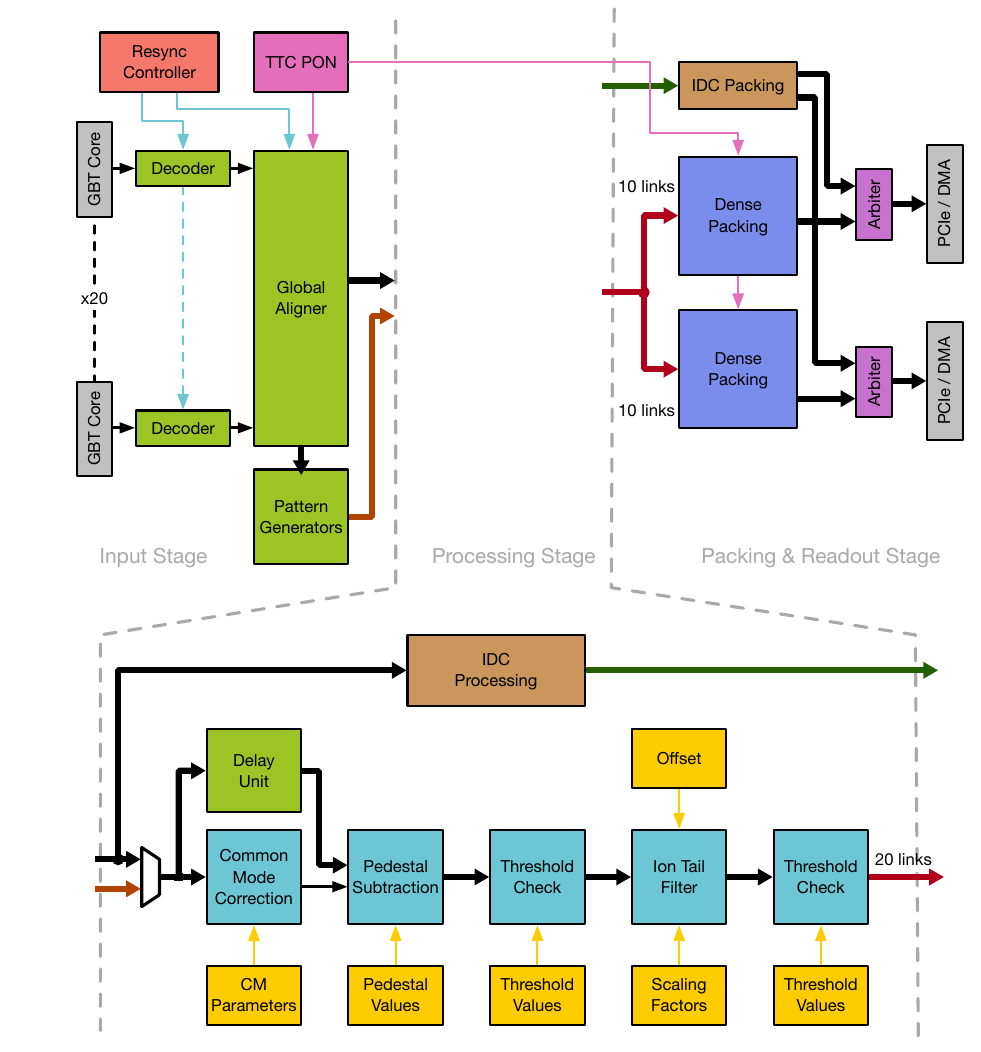}
  \caption{Block diagram of the \gls{TPCUL}.
  }
  \label{fig:UL:Overview}
\end{figure}

The \gls{UL} design can be functionally divided into three distinct stages: 
\emph{\textit{input stage}}, \emph{\textit{processing stage}}, and \emph{\textit{output stage}}.
A block diagram is shown in \figref{fig:UL:Overview}.

\begin{itemize}
    \item The input stage decodes the data from the \gls{GBT} frames and generates the link data streams together with the time info stream.
    It also handles stream synchronization and alignment.
    A detailed description is given in \secref{sec:UL:Decode}.
    \item The processing stage applies the necessary corrections to ensure the desired data quality.
    It also performs suppression of samples below threshold to reduce the output data volume.
    Details are provided in \secref{sec:UL:Blocks}.
    \item The output stage receives the link data streams and the time info stream, extracts the processed samples, and packs them into a dense format for transmission as data blocks.
    Details are provided in \secref{sec:UL:Packing}.
\end{itemize}

\subsection{Input stage}
\label{sec:UL:Decode}

The primary responsibilities of the \textit{input stage} are the decoding, synchronization, and alignment of the incoming data.
Data are received by the \gls{CRU} Common Logic and forwarded to the \gls{UL} in the form of \gls{GBT} frames, whose payload is detector-specific.
For the \gls{TPC}, eight \gls{GBT} frames carry the 80 \gls{ADC} samples corresponding to one \gls{ADC} sampling clock cycle (i.e., one time-bin), acquired from two and a half SAMPA chips (\cf~Sections~\ref{sec:TPC:FEC} and \ref{sec:TPC:GBTframes}).
 
\subsubsection{GBT frame decoder}
\label{sec:UL:Decode:Decode}

The data format of the \gls{GBT} frames was described in \secref{sec:TPC:GBTframes}.
For each optical input link, a dedicated \gls{GFD} extracts the samples from the \gls{GBT} frames.

After initialization or reset, the \gls{GFD} searches for the \textit{SYNC} pattern to align with the start of the \gls{ADC} data.
Once locked, the samples can be assigned to channels according to their deterministic position in the stream, which cycles through the \num{16} channels time-bin by time-bin, with the first samples after \textit{SYNC} corresponding to channels 0 or 16 (depending on the SAMPA stream).

The decoded samples are forwarded to the aligner (\cf~\secref{sec:UL:Decode:Align}) as an intermediate stream with two samples per \gls{CRU} clock cycle.
This stream is not yet the final link data stream, as alignment and timing information are still missing.

In parallel, the stream carries the \SI{5}{MHz} \gls{ADC} sampling clock oversampled by the \SI{160}{MHz} \gls{ELink} clock (\cf~\secref{sec:TPC:FEC}).
The decoder monitors these clocks to verify phase alignment across all SAMPAs in the \gls{TPC}.

\subsubsection{Global aligner}
\label{sec:UL:Decode:Align}

\begin{figure}[ht]
  \centering
  \includegraphics[width=0.99\linewidth]{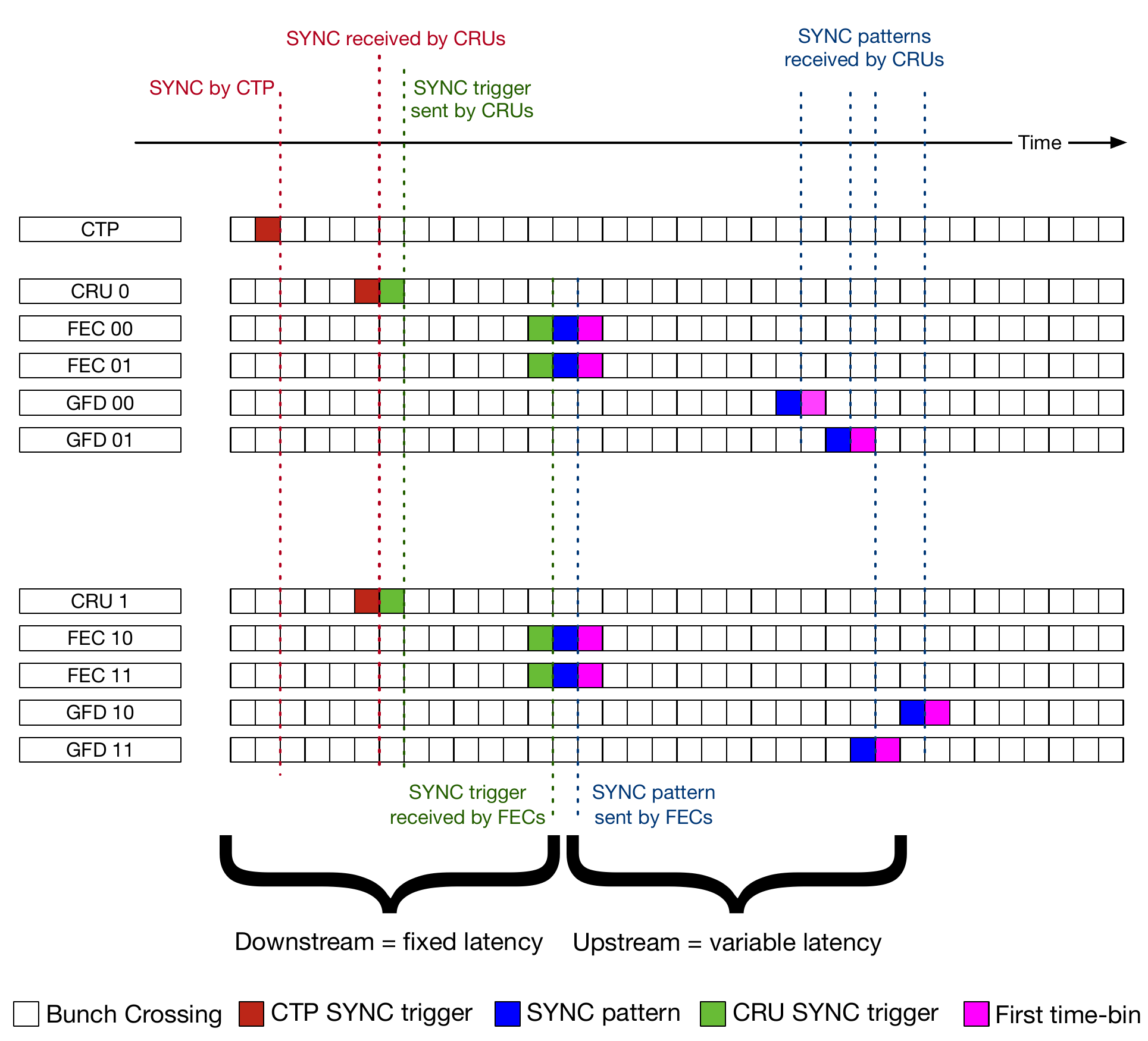}
  \caption{Diagram illustrating the synchronization timeline of the \gls{FEE}.
  For clarity, only two \glspl{CRU} are shown, each connected to two \glspl{FEC}.
  Within the \gls{CRU}, the incoming \gls{GBT} link data are decoded by the \gls{GFD} module.}
  \label{fig:ALIGNMENT}
\end{figure}

Since the Tx links from the \glspl{FEC} to the \glspl{CRU} have variable latency, the \textit{SYNC} patterns may arrive at different times on different links (\cf \figref{fig:ALIGNMENT}).
However, samples from the same time-bin must be processed together (\cf~\secref{sec:UL:Dataformat}).
To ensure temporal alignment before entering the processing pipeline, a global aligner module buffers and synchronizes the input streams.
The aligner exploits the ordered data structure and the known position of the \textit{SYNC} pattern.
Each decoded input stream (with the SYNC pattern removed and starting at time-bin 0) is written into a dedicated FIFO, which is reset before buffering begins.
At a predefined reference time $t_\text{align}$, all FIFOs are read out simultaneously.
The reference time is chosen to exceed the maximum expected round-trip latency, ensuring that all FIFOs contain valid data before readout begins; larger values are possible if sufficient FIFO depth is available.
The global aligner outputs the link data streams and the time info stream described in \secref{sec:UL:Dataformat}.
It also generates control and synchronization flags such as \textit{TB Start}, \textit{First TB}, and \textit{Resync}.
These streams are then forwarded to the processing stage (\cf \secref{sec:UL:Blocks}).

\subsubsection{Resync controller}
\label{sec:UL:Decode:Resync}

A dedicated resync controller enables re-synchronization in case of alignment loss during data taking.
This may happen for example in the case of a link loss due to radiation.
The following describes the mechanism used to re-synchronize the data streams at any moment.

A re-synchronization request from the Pattern Player (\cf~\secref{sec:TPC:Synchronisation}) in the \gls{CRU} Common Logic is distributed to the \glspl{FEC} and the resync controller.
After a programmable delay $t_\text{wait}$, chosen slightly below the minimum round-trip time (to limit data loss), the controller resets the \gls{GFD} modules.
The decoders then halt transmission and wait for the next \textit{SYNC} pattern.

During this period, the global aligner continues readout.
To maintain a continuous and well-formed output stream, valid timing information is preserved while sample values are temporarily replaced with zeros, allowing the FIFOs to be reset safely.

Once the \gls{GFD} modules re-lock to the \textit{SYNC} pattern, the aligner buffers the new data and resumes normal operation after the programmable delay $t_\text{align}$.

Throughout the procedure, the link data and timing streams remain continuous and correctly formatted, ensuring transparent behavior for downstream processing.

\paragraph{Performance}

During \gls{TPC} operation with beam in 2022, radiation-induced link losses were observed.
The effect exhibited a clear radial dependence (fewer losses at larger radii) and scaled with interaction rate.
At the target \PbPb interaction rate of \SI{50}{\kHz} and without mitigation, the projected acceptance loss of 5–10\,\% per hour would be unacceptable for physics data taking.

\begin{figure}[ht]
  \centering
  \includegraphics[width=0.7\linewidth]{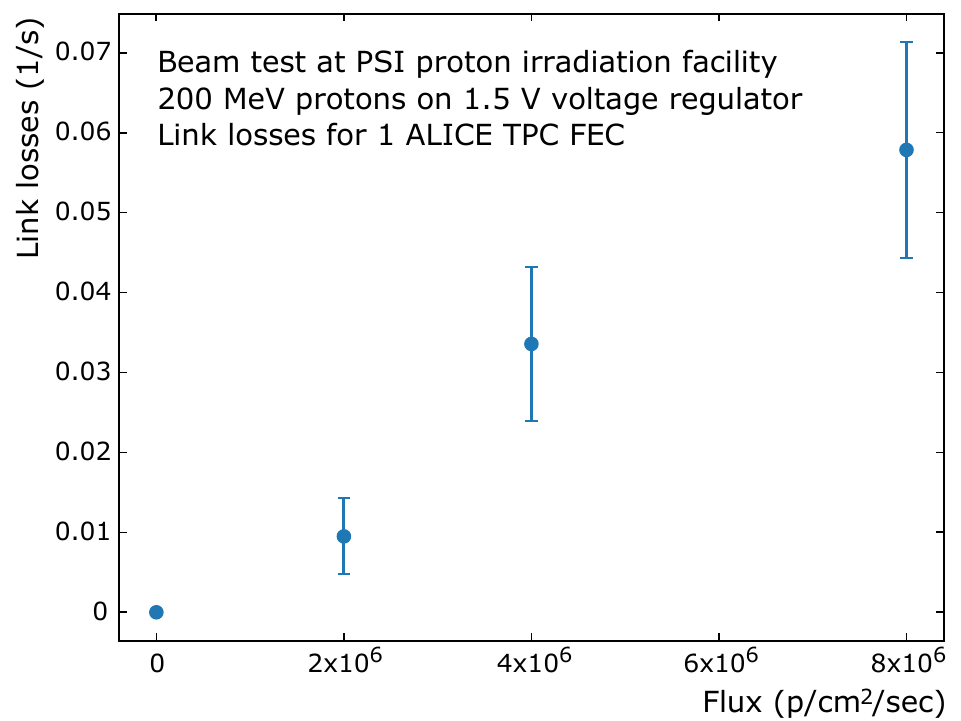}
  \caption{Link losses as a function of beam intensity in a test at PSI with \SI{200}{\MeV} proton beams at PSI.
  The beam was focused on the the \SI{1.5}{\V} voltage regulator on the \gls{TPC} \gls{FEC}.}
  \label{fig:TPC_Linklosses}
\end{figure}

The origin of the issue was identified in dedicated irradiation tests at the Proton Irradiation Facility (PIF) at Paul Scherrer Institut (PSI) using \SI{200}{\MeV} protons.
The measured link loss rate as a function of beam intensity is shown in \figref{fig:TPC_Linklosses}.
The effect was traced to radiation-induced perturbations of the \SI{1.5}{\V} regulator supplying the Gigabit Transceivers and associated digital circuitry, most likely affecting the \gls{GBTx} PLL.

As the \gls{FEE} was already installed, the issue could not be removed but had to be mitigated.
The link losses are temporary and do not affect \gls{FEC} functionality; however, synchronization at the \gls{CRU} is no longer guaranteed after a link re-establishes.
Therefore, in routine operation periodic re-synchronization is performed every 10 \glspl{TF}, corresponding to a rate of \SI{35}{\hertz}.
The re-synchronization is performed near the end of the \gls{TF}, ensuring stable and transparent data taking.

\subsubsection{Pattern generator}
\label{sec:UL:Pattern}

The pattern generator produces a deterministic test stream that traverses the entire \gls{UL} processing chain, enabling functional validation, particularly of the output stage (\cf \secref{sec:UL:Packing}).
It uses a \nbit{32} linear feedback shift register (LFSR) with configurable threshold and selectable fixed-pattern modes (constant, channel ID, time-bin, or a combination) to emulate controlled occupancies and embed verifiable markers into each link data stream with minimal architectural overhead.

\subsection{Processing stage}
\label{sec:UL:Blocks}

As shown in \figref{fig:UL:Overview}, the \gls{TPCUL} processing stage consists of different blocks dedicated to specific tasks in the data pipeline.
These blocks operate in real-time, applying corrections and filtering as data streams through the system.
The following subsections describe each block’s function, implementation, and role within the overall architecture.

\subsubsection{Common-mode correction}
\label{sec:UL:Blocks:CMC}

A \emph{\textit{common-mode effect}} refers to a baseline shift shared among several (or a group of) detector channels.
In the \gls{TPC} \gls{GEM} system, this effect originates from capacitive coupling between the pad plane and the adjacent GEM foil.
It causes all pads beneath a given \gls{GEM} stack to exhibit a common, positive or negative baseline displacement in the presence of signals on the pad plane.

The common-mode signal is characterized by a bipolar waveform: A negative peak coinciding with the induced (positive) signal, followed by an overshoot and a long positive tail.
The tail is approximately exponential, with a time constant defined by the RC circuit formed by the \gls{GEM} electrode’s loading resistor and its capacitance to ground (including the HV cable).
While the negative and positive components of the waveform cancel on average under constant occupancy, fluctuations in the signal rate may break this balance.
For instance, high-occupancy events---\eg central heavy-ion collisions---result in positive baseline shifts in subsequent time periods with lower occupancy due to the lingering tails.
On the other hand, large energy deposits in the Landau tail can create negative shifts that will not be balanced by prior positive tails.

The baseline variations due to common-mode effect impact zero suppression.
Positive shifts may allow noise to pass, increasing data volume and reducing signal fidelity.
Negative shifts, on the other hand, reduce---or may even suppress---genuine signals, leading to information loss.
Therefore, correcting the common-mode effect prior to zero suppression is essential for maintaining the tracking and particle identification performance of the \gls{TPC}.

A detailed description and discussion of the correction method can be found in \cite{CMCITpaper}.
The chosen correction method is the following:
The common-mode value is estimated for each time-bin using pads without any physical signal (referred to here as \textit{\emph{empty pads}}).
The value is then subtracted from all samples in the given time-bin.

As shown in \figref{fig:UL:Overview}, the common-mode correction module is placed upstream of the pedestal subtraction stage.
During the computation of the common-mode value, the link data streams and the time-info stream are delayed to maintain temporal alignment.
Once computed, the common-mode value is passed to the pedestal subtraction module (\cf~\secref{sec:UL:Blocks:PedZS}), where it is added to the samples concurrently with the pedestal subtraction.

To perform the calculation of the common-mode value, the pedestal values must also be subtracted from all samples.
The pedestal values are available within the common-mode correction module, as they are included in the link data streams (\cf~\secref{sec:UL:Dataformat}).

\paragraph{Pad-by-pad scaling}

The capacitive coupling between the pad plane and the adjacent \gls{GEM} foil is not uniform across the detector.
Local variations arise from geometrical imperfections, such as wrinkles caused by imperfect foil stretching.
These non-uniformities modulate the amplitude of the common-mode signal induced on different pads.

To quantify this effect, dedicated calibration-pulser runs are used.
In these runs, a voltage pulse with known amplitude and timing is injected into the \gls{GEM} electrode facing the pad plane, inducing signals on all pads beneath it (\cf~\cite{TPCjinst}).
The measured pulser response reflects the local capacitive coupling, and the normalized pulser charge $k_\text{pad}$ provides a relative measure of the per-pad coupling strength.
This parameter is used to scale the common-mode correction individually for each pad.
Both $k_\text{pad}$ and its inverse $1/k_\text{pad}$ are precomputed and supplied to the \gls{UL} as configuration parameters%
\footnote{Divisions are more resource-intensive than multiplications in \glspl{FPGA}; providing both values avoids real-time division and thus improves performance.}%
, ensuring efficient implementation in the \gls{FPGA}.

\paragraph{Algorithm overview}

The common-mode (CM) effect affects all pads under the same GEM stack in a similar manner, independent of whether a physical signal is present. Pads without signal contain only CM and electronic noise and can therefore provide the basis for estimating the CM contribution for each time-bin.

Let $q_\text{pad}$ denote the measured sample value. The correction algorithm proceeds in two stages:

\begin{enumerate}
\item \textit{Empty-pad identification and CM estimation:}
Candidate empty pads are first selected using a threshold criterion
    \begin{equation}
        \frac{q_\text{pad}}{k_\text{pad}} < T_1 \ ,
        \label{sec:UL:Blocks:cmcthr}
    \end{equation}
where $k_\text{pad}$ is the per-pad correction factor and $T_1$ is a configurable threshold.
Since residual signal tails from previous time-bins may still contribute below this threshold, an additional statistical consistency check is applied.
For each candidate pad, the scaled sample is compared to ten randomly selected pads $B_n$, likewise scaled:
\begin{equation}
    \left\lvert \frac{q_\text{pad}}{k_\text{pad}} - \frac{q_{(B_n)}}{k_{(B_n)}} \right\rvert < d_\text{match} \, .
    \label{sec:UL:Blocks:cmcmatch} 
\end{equation}
If this condition is satisfied more than $N$ times, where $N$ is a configurable parameter, the pad is confirmed as empty.  

The CM value for the given time-bin is then calculated as the average over all confirmed empty pads.

\item \textit{Pad-wise correction:}  
The corrected sample is obtained as
\begin{equation}
    q_\text{pad}^{\text{corr}} = q_\text{pad} - k_\text{pad} \cdot \text{CM} \, ,
    \label{sec:UL:Blocks:cmcorr}
\end{equation}
where the scaling factor $k_\text{pad}$ again accounts for local variations in capacitive coupling.
\end{enumerate}

As shown in \cite{CMCITpaper}, the combined threshold and statistical matching procedure provides a stable and unbiased CM estimate even in the presence of overlapping signal tails, enabling reliable pad-by-pad correction tailored to the local detector response.

\paragraph{Implementation}

Although the common-mode correction algorithm is conceptually straightforward in software, its implementation in firmware is considerably more demanding due to stringent timing requirements and the need for extensive loop unrolling and parallel execution.

\begin{figure}[htb]
    \centering
    \includegraphics[width=\linewidth]{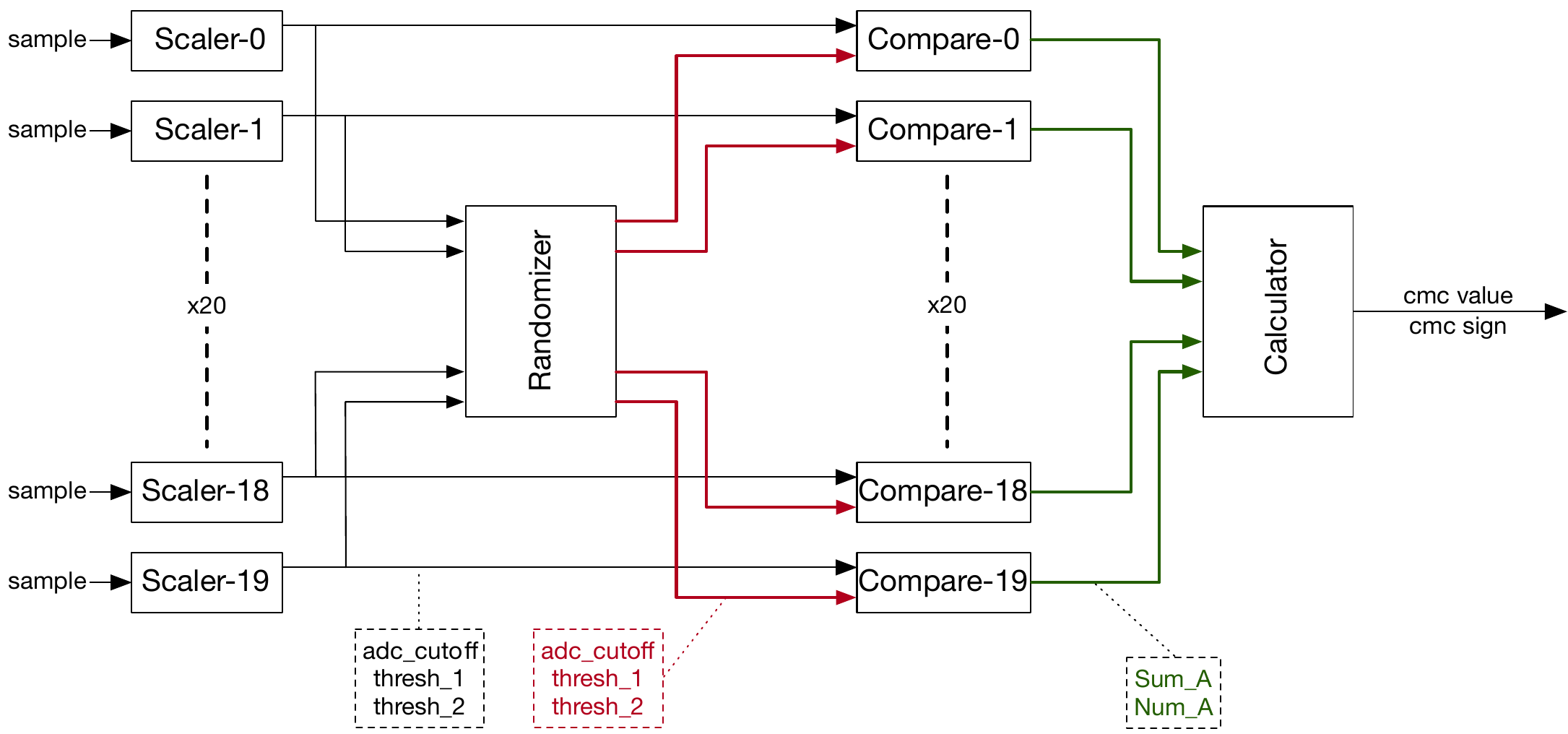}
    \caption{Top level overview of the common-mode correction module, consisting of \num{20} scaler units (one per link), one randomizer unit, \num{20} compare and match units and the CM calculator.
    The output of each scaler unit are the clipped sample value and the two flags {\tt thresh\_1} and {\tt thresh\_2}.
    The output of the randomizer unit are \num{10} clipped sample values and the two flags {\tt thresh\_1} and {\tt thresh\_2} each for each compare unit.}
    \label{fig:cmc:top}
\end{figure}

A top-level overview of the design is shown in \figref{fig:cmc:top}.
It comprises, for each link data stream, a scaler unit and a compare-and-match unit, as well as a global randomizer unit and the common-mode calculator.
The system processes two samples per link data stream and clock cycle, corresponding to \num{40} samples per cycle across the \num{20} input links (\cf~\secref{sec:UL:Dataformat}).
The comparison stage constitutes the primary performance bottleneck:
$40$\ samples\ $\times\ 10$\ comparisons\ $= 400$ parallel comparators.
This large number of comparators dominates the resource usage and directly constrains scalability.

The numerical representation of the input samples is a critical factor in comparator size and speed.
A standard \nbit{32} floating-point format would require large, slow comparators and additional pipelining, further increasing resource consumption.
To overcome this, the design employs a custom fixed-point representation optimized for the specific needs of each submodule.
Both dynamic range and precision are tuned locally to achieve the required accuracy while minimizing hardware usage.

\begin{figure}[htb]
    \centering
    \includegraphics[width=.95\linewidth]{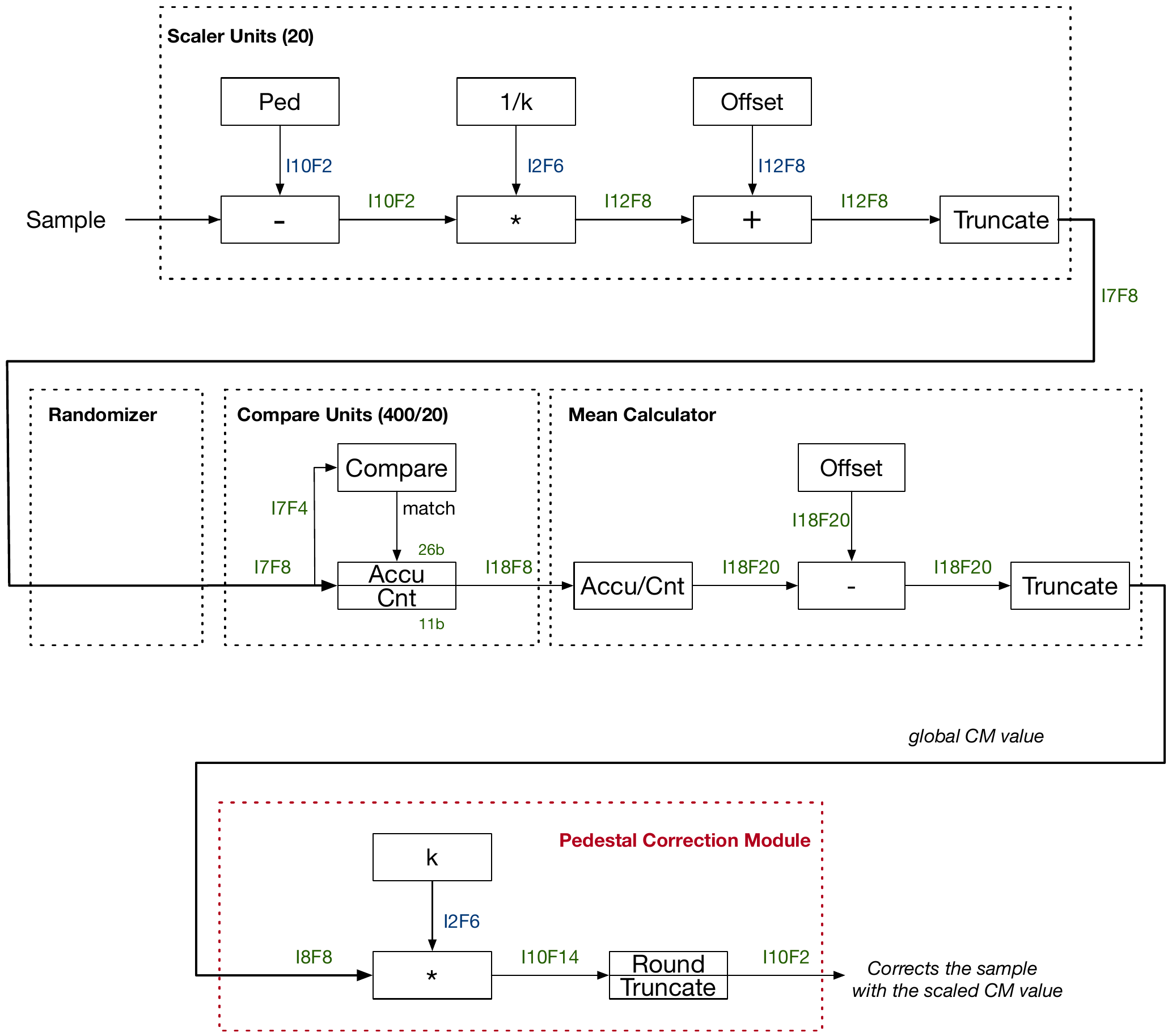}
    \caption{Associated fixed-point bit widths for the different arithmetic and logic stages of the common-mode correction pipeline.
    All calculations use a fixed-point format I$x$F$y$, where $x$ is the number of bits used for the integer part and $y$ the number of fractional bits.}%
    \label{fig:cmc:calc}%
\end{figure}

\figref{fig:cmc:calc} provides a detailed view of the modules implementing the individual operations, including the fixed-point precision chosen for each submodule.
These components are discussed in the following sections.

The \textit{ADC valid} signal (\cf~\secref{sec:UL:Dataformat}) is propagated through a dedicated pipeline and used to control the accumulators in the compare-and-match units as well as in the common-mode calculator, ensuring that only valid samples are used.
For clarity, this control path is not shown in the corresponding figures.

\subparagraph{Scaler units}

\begin{figure}[htb]
    \centering
    \includegraphics[width=.9\linewidth]{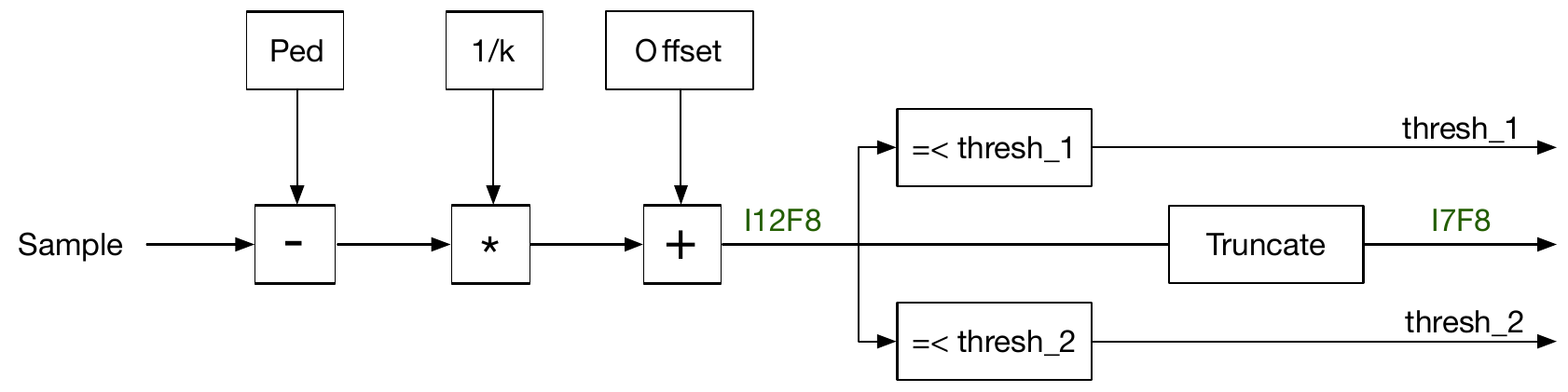}
    \caption{Operations of the scaler unit.
    The necessary threshold checks are performed at this stage in order to reduce the complexity of the subsequent compare units.}%
    \label{fig:cmc:scaler}%
\end{figure}

The scaler units serve as the pre-processing stage for the calculation of the common-mode value.
The arithmetic operations are detailed in \figref{fig:cmc:scaler}.

For each sample, three primary operations are applied:
(1) subtraction of the corresponding pedestal value,
(2) application of the inverse pad-specific gain factor $1/k_\text{pad}$, and
(3) addition of a fixed offset.
The purpose of this offset is to shift the baseline into the positive domain, as the common-mode effect may result in a strong negative baseline displacement.
By ensuring that all intermediate values are positive, subsequent processing is simplified.

The scaled values are truncated to comply with the limited range of the common-mode correction algorithm, from \num{-100} to +28, and are encoded using \SI{7}{bits}.
Consequently, any result exceeding $T_2=28$ is clipped, and the corresponding sample is flagged using the \variable{thresh\_2} indicator to denote that it falls outside the valid range.

Additionally, the scaler units perform the evaluation of the empty-pad pre-selection criterion \eqref{sec:UL:Blocks:cmcthr} by comparing the scaled charge to the threshold $T_1$ (\variable{thresh\_1} in \figref{fig:cmc:scaler}).
Samples below the threshold are flagged using the \variable{thresh\_1} indicator.
Performing this check at the scaler stage reduces logic complexity and minimizes the number and bit-width of comparators required in the subsequent comparator stage, where ten units operate in parallel.

\subparagraph{Randomizer unit}

The randomizer unit interfaces between the \num{20} scaler units and the \num{20} comparator units.
It accepts \num{20} input data streams and produces \num{10} distinct output streams per comparator unit, a total of \num{200} output streams.
Despite its name, the randomizer does not perform stochastic or dynamic mapping; rather, it employs a predetermined, fixed selection pattern.

Initially, the architecture was conceived to incorporate true random selection of input streams for each comparator unit.
However, this approach was ultimately abandoned due to the excessive logic resource demands associated with implementing randomized selection in hardware.
Moreover, the spatial distribution of pads across the scaler units is inherently randomized to a sufficient degree, rendering explicit randomization redundant.

To validate the adequacy of the fixed mapping strategy, a software reference model employing randomized input selection was compared against the fixed-hardware configuration.
No statistically significant deviation in performance or behavior was observed, thereby justifying the use of the fixed mapping in the hardware implementation.

\subparagraph{Compare and match units}

\begin{figure}[ht]
    \centering
    \includegraphics[width=.93\linewidth]{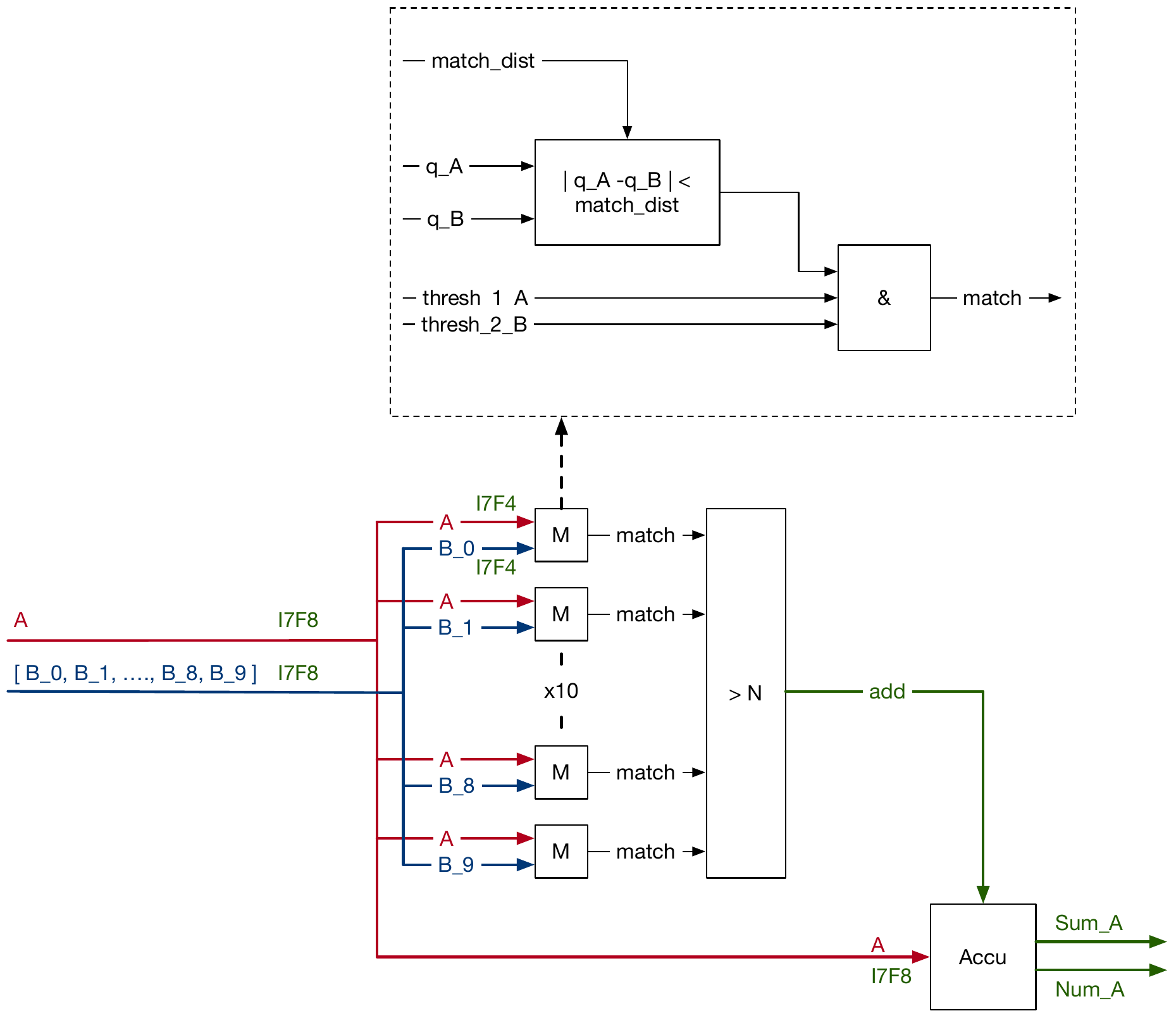}
    \caption{Compare unit with \num{10} parallel match units.
    Matching is performed at reduced precision (I7F4) to conserve resources, while accumulation is carried out at standard precision (I7F8).
    The accumulator sums the sample values for all confirmed empty pads (\variable{Sum\_A}) and counts the number of empty pads (\variable{Num\_A}) for the given time-bin and for the given link data stream.}
    \label{fig:cmc:comp}%
\end{figure}

\Figref{fig:cmc:comp} shows the internal structure of the compare unit, which includes ten match units.
The compare unit performs a parallel comparison between an incoming sample $q_A$ and a set of ten randomly selected reference samples $q_{(B_n)}$, $n \in \{0\ldots 9\}$.
Each comparison is handled by a dedicated match unit operating concurrently.

Each match unit checks whether the matching condition \eqref{sec:UL:Blocks:cmcmatch} is fulfilled (the values are already scaled using the per-pad correction factors).
To optimize resource usage, the comparison ignores the four least significant bits of the fixed-point representation, as they have negligible impact on the result.

In addition to the amplitude comparison, each match unit enforces two auxiliary conditions:
The reference value $q_{(B_n)}$ must lie within the valid common-mode correction range and must not be clipped.
This is indicated by the \variable{thresh\_2\_B} flag received from the randomizer unit.
The candidate value $q_A$ must also satisfy the pre-selection criterion \eqref{sec:UL:Blocks:cmcthr}, indicated by the \variable{thresh\_1\_A} flag.
A match is registered only if all three conditions are simultaneously fulfilled.%
\footnote{The parameters {\tt thresh\_2\_A} and {\tt thresh\_1\_B} are not utilized within this specific match unit; however, in a complementary unit the roles of A and B are interchanged. To ensure a uniform signal pipeline architecture, all four threshold values are consistently supplied regardless of their local usage.}

If the number of matches exceeds a configurable threshold $N$, the sample $q_A$ is classified as originating from an empty pad.
It is then included in the calculation of the common-mode value.

An accumulator module keeps track of both the number of valid matches and the sum of the corresponding sample values for each link data stream and time-bin.

\subparagraph{Common-mode calculator}

\begin{figure}[ht]
    \centering
    \includegraphics[width=.95\linewidth]{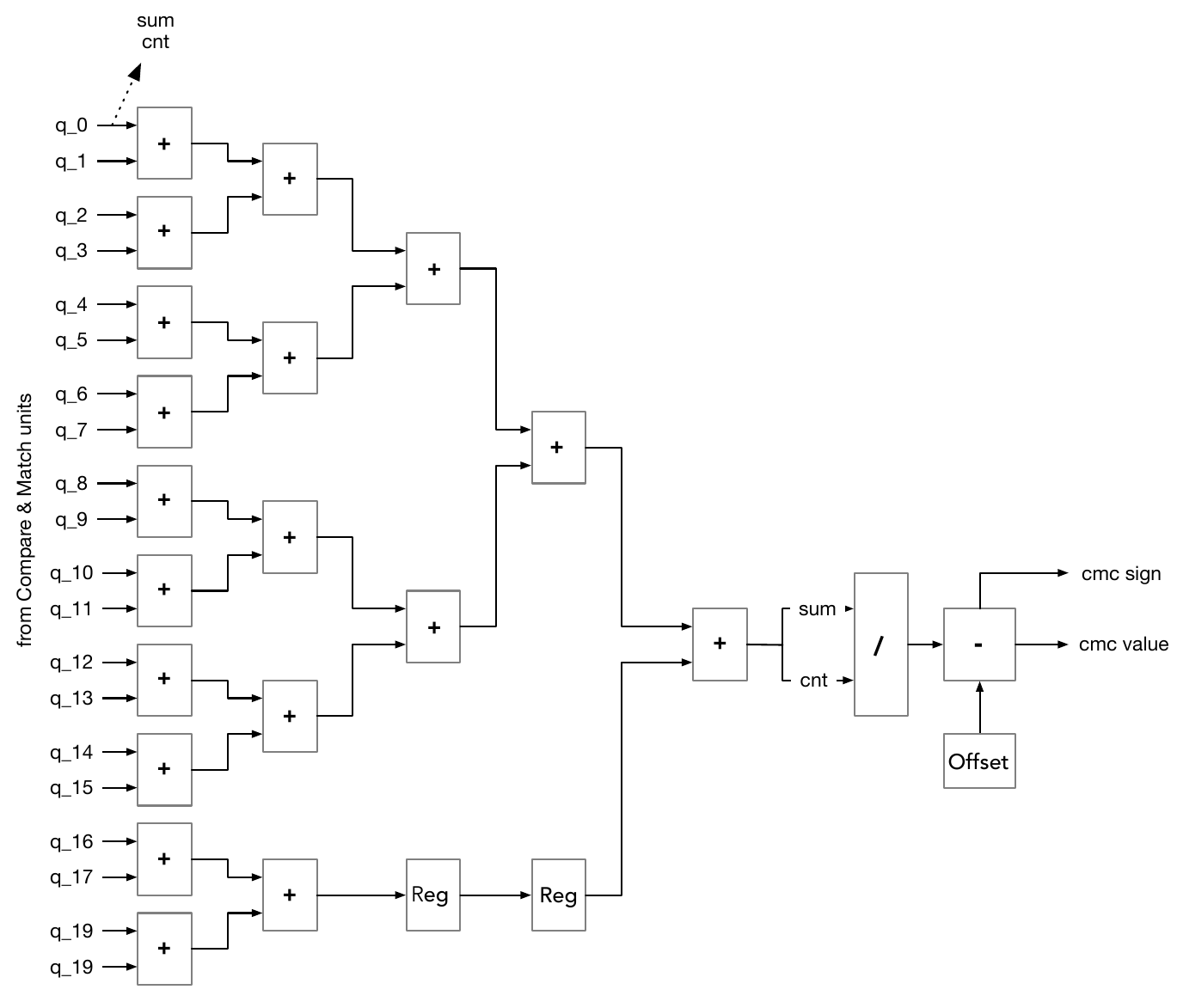}
    \caption{The common-mode calculator aggregates the partial sums and match counts from all compare and match units using a five-stage pipelined adder tree.
    A fixed-point divider then computes the mean, which is subsequently corrected by subtracting the offset introduced by the scaler units, yielding the final common-mode value.}%
    \label{fig:cmc:calculator}%
\end{figure}

The common-mode calculator (\cf \figref{fig:cmc:calculator}) performs the final computation of the global common-mode value for a given time-bin.
It evaluates the mean over all identified empty pads, as determined by the preceding compare and match units.

Due to the high resource and latency cost of division operations in \gls{FPGA} implementations, division is intentionally deferred to the final stage.
Earlier stages, specifically the compare and match units, output the accumulated charge sum and the corresponding count of empty pads, rather than mean values.

The calculator module aggregates these partial sums and counts from all \num{20} comparator units to obtain the total accumulated charge and the total number of contributing pads.
To meet the stringent \SI{240}{MHz} timing requirement, this aggregation is performed using a five-stage pipelined adder tree.
Each stage consists of registered adders that process intermediate results from the previous stage.
Since the tree’s full capacity of \num{32} inputs ($2^5$) is not fully utilized, the structure is asymmetric, and some paths contain pass-through registers instead of arithmetic units.

The outputs of the adder tree (total sum and total count) are subsequently processed by a fully pipelined fixed-point divider.
This unit provides a new result every clock cycle after an initial latency of eight cycles.%
\footnote{This is the latency of the divider itself.}
Finally, the fixed offset introduced earlier by the scaler units is subtracted from the result to yield the true common-mode value.
The output consists of the unsigned common-mode magnitude, along with a sign flag indicating the direction of the baseline shift.
A valid output is calculated every \num{48} clock cycles, when all the samples in a time-bin have been processed.
This information is then forwarded to the pedestal subtraction and zero-suppression modules (\cf \secref{sec:UL:Blocks:PedZS}), where the actual sample corrections are performed following \eqref{sec:UL:Blocks:cmcorr}.

\paragraph{Delay unit}

A delay unit (\cf \figref{fig:UL:Overview}), operating in parallel with the common-mode correction module, compensates for its processing latency by delaying the link data streams and the time-info stream.
This ensures that the raw samples and the computed global common-mode value reach the pedestal-subtraction and zero-suppression modules in the same clock cycle.
Synchronization is maintained using compact embedded FIFOs.

\begin{figure}[ht]
    \centering
    \includegraphics[width=.99\linewidth]{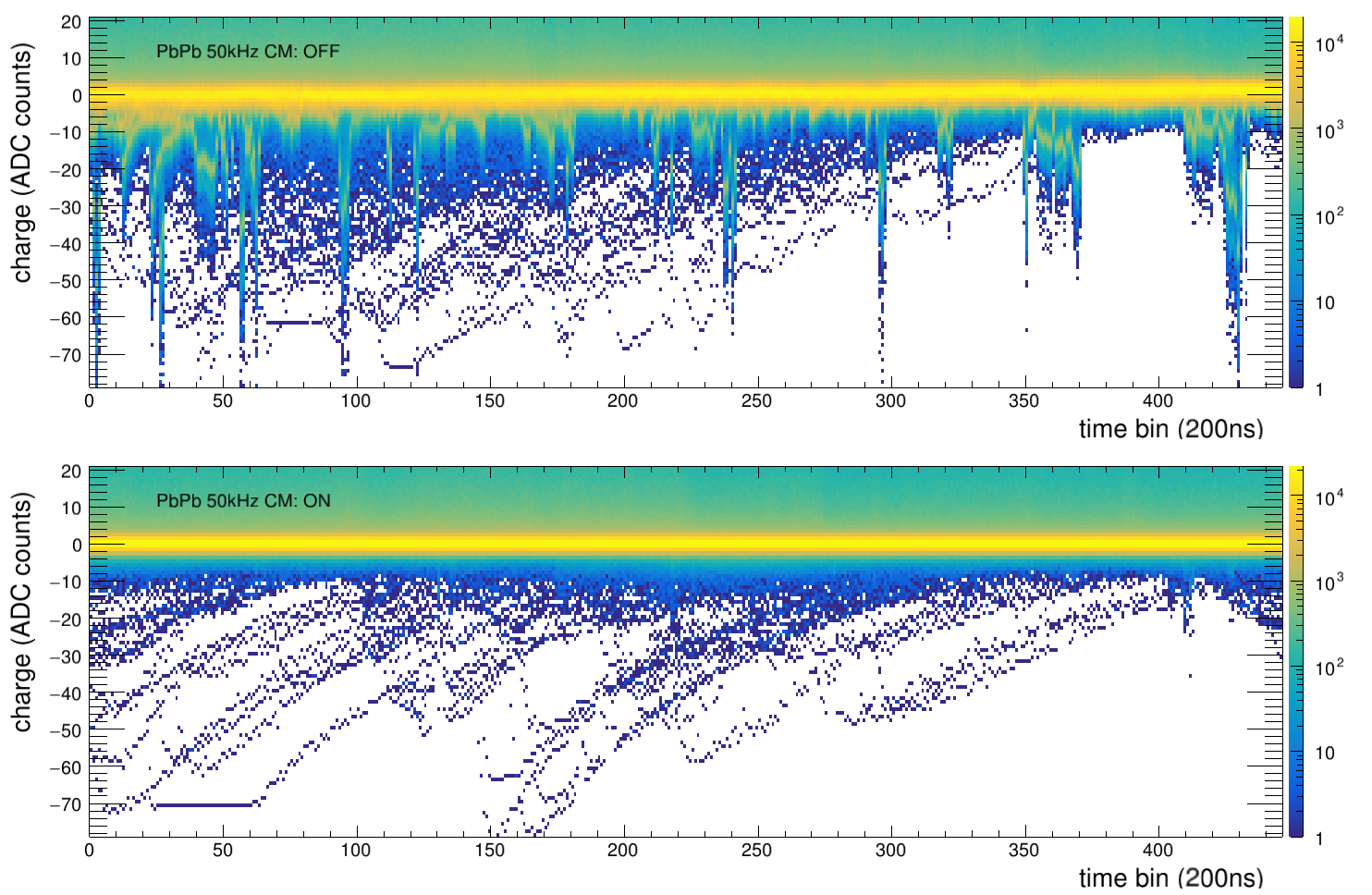}
        \caption{Overlay of the data from all channels connected to one \gls{CRU} and from many \glsps{HBF}, with common-mode correction disabled (CM OFF, upper panel) and enabled (CM ON, lower panel).
        The data were recorded in two different runs with \PbPb collisions at \SI{50}{\kilo\hertz}.}%
    \label{fig:cmc:example}%
\end{figure}

\paragraph{Performance}

The common-mode correction algorithm was validated on the \gls{TPC} \glspl{CRU} using \PbPb collisions at \SI{50}{\kilo\hertz}.
To enable detailed analysis, zero suppression was disabled, allowing the full raw data to be transmitted off-detector.
Triggered readout (with a random trigger) was employed in order to reduce the data rate.
Multiple data sets were acquired under varying configurations of the common-mode correction module for comparison.%
\footnote{The figures in this paper are created using the optimum configuration that was found in these tests: threshold $T_1= 12$, match distance $d_\text{match}=3$, and number of matches needed to confirm an empty pad $N=5$.}

\Figref{fig:cmc:example} shows an overlay of the data from all channels connected to a single \gls{CRU}, across many \glsps{HBF}, with common-mode correction disabled (top) and enabled (bottom).
The baseline, expected at 0 \gls{ADC} counts, is clearly visible in both cases.
Without correction, numerous short outliers toward negative \gls{ADC} values are observed. These correspond to common-mode signals and can reach amplitudes of several tens of \gls{ADC} counts.
With correction enabled, these signals are effectively suppressed, and the baseline remains highly stable.
The few remaining negative excursions are localized to individual channels and are unrelated to the common-mode effect.

\begin{figure}[ht]
    \centering
    \includegraphics[width=.5\linewidth]{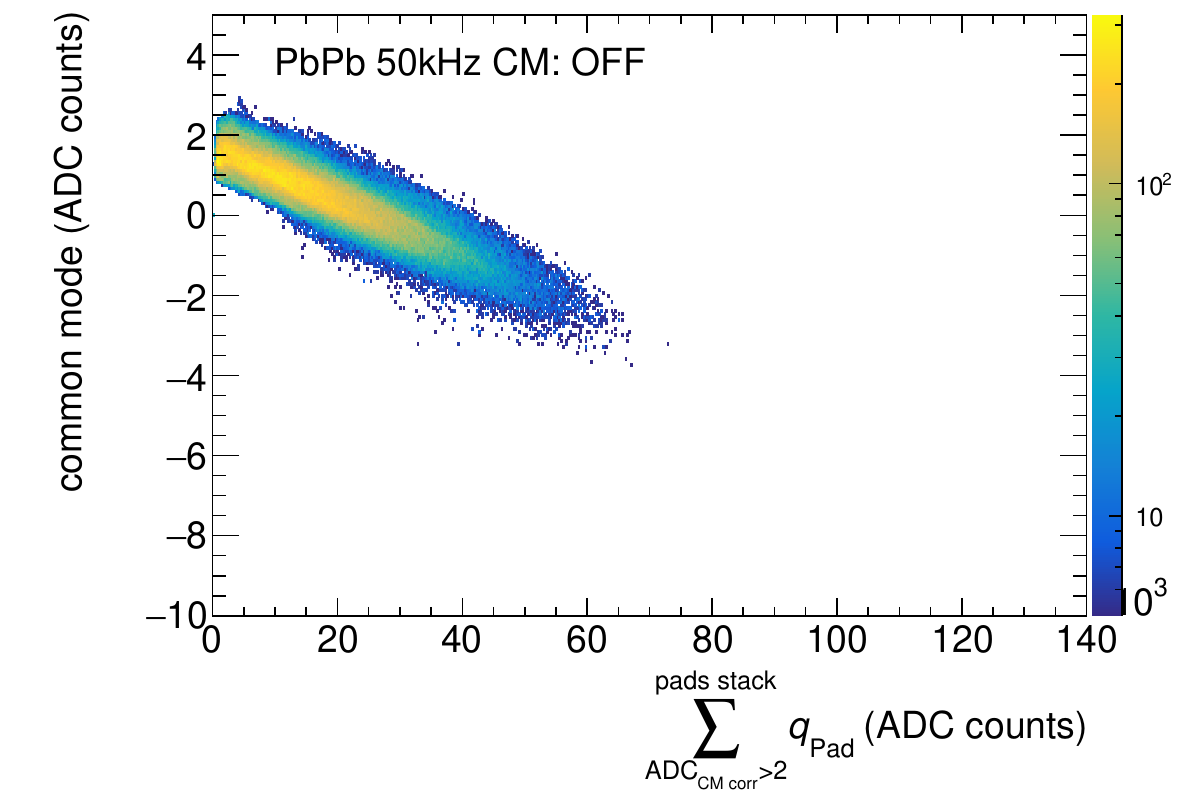}\hfill
    \includegraphics[width=.5\linewidth]{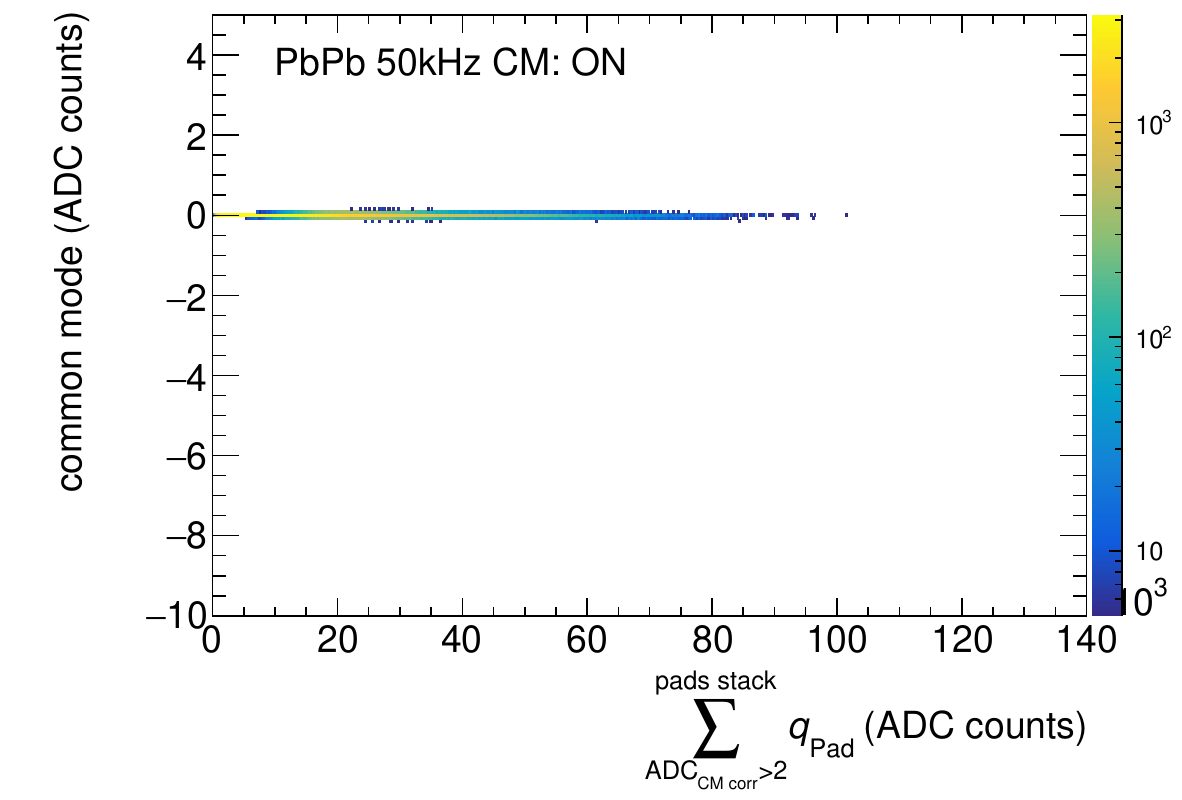}
    \caption{Correlation between the sum of all positive signals found in a given time-bin in one \gls{CRU} and the calculated common-mode value, with common-mode correction disabled (CM OFF, left panel) and enabled (CM ON, right panel).
    The data were recorded in two different runs with \PbPb collisions at \SI{50}{\kilo\hertz}.}%
    \label{fig:cmc:correlate}%
\end{figure}

\Figref{fig:cmc:correlate} further illustrates the nature of the common-mode signal and the effectiveness of the correction.
It plots the correlation between the positive induced signals in a given time-bin and the calculated common-mode value in the same time-bin.
When correction is disabled, the expected anti-correlation is observed: large positive signals coincide with increasingly negative common-mode values.
The figure also confirms that positive common-mode values can occur, highlighting the bipolar nature of the effect.
With correction enabled, the described dependence disappears, confirming the removal of all (negative and positive) common-mode contributions.

\begin{figure}[ht]
    \centering
    \includegraphics[width=.5\linewidth]{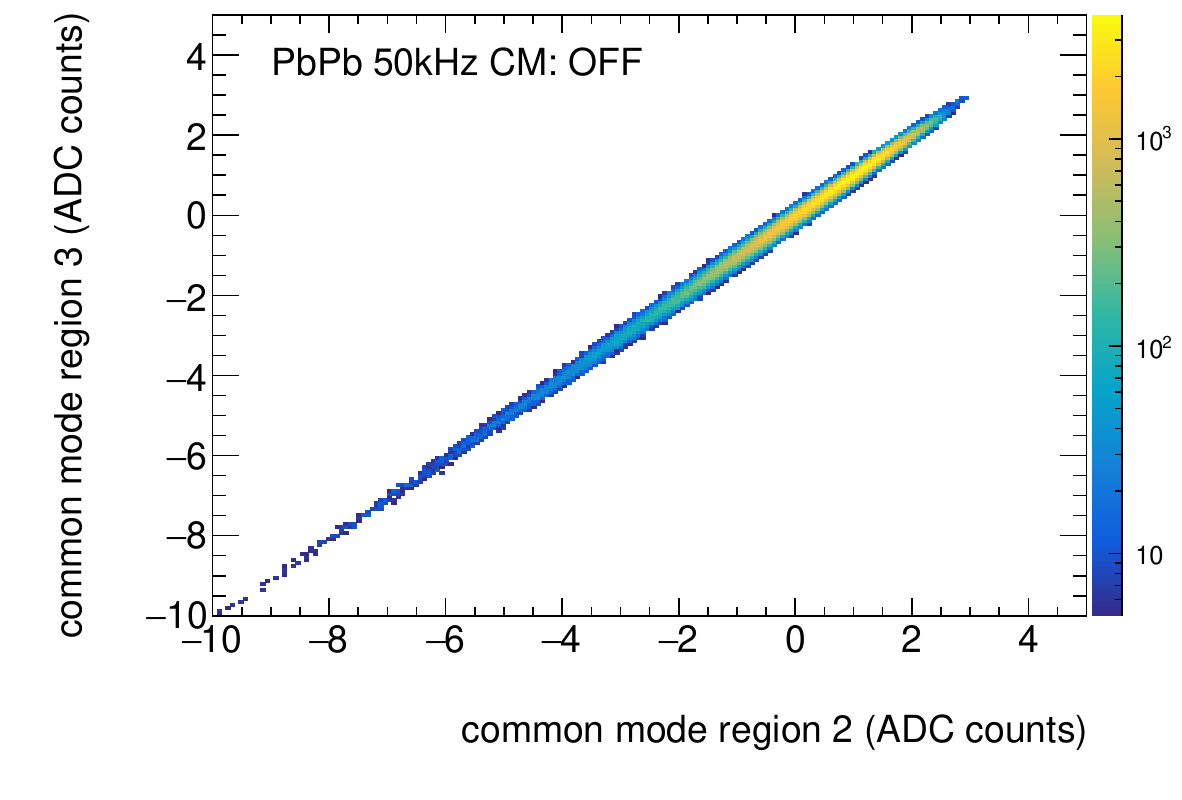}\hfill
    \includegraphics[width=.5\linewidth]{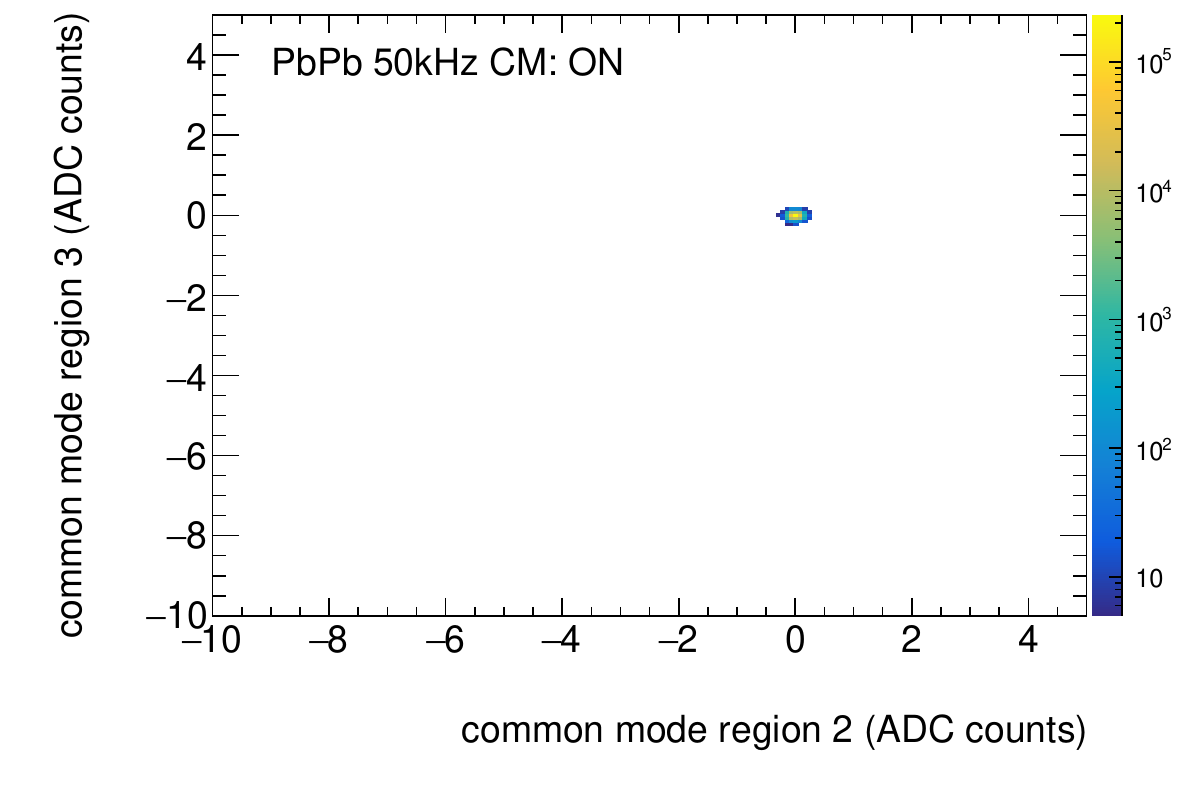}
    \caption{Correlation between the calculated common mode values for two \glspl{CRU} covering two pad regions under the same \gls{GEM} stack, with common-mode correction disabled (CM OFF, left panel) and common-mode correction enabled (CM ON, right panel).
    The data were recorded in two different runs with \PbPb collisions at \SI{50}{\kilo\hertz}.}%
    \label{fig:cmc:compare}%
\end{figure}

Since all pads under the same \gls{GEM} stack are similarly affected by common-mode shifts, and since multiple \glspl{CRU} may read out pads from the same stack, the consistency of the correction can be further verified.
\Figref{fig:cmc:compare} shows the correlation between the calculated common-mode values from two different \glspl{CRU} reading out data from pads under the same \gls{GEM} stack.
With correction disabled, a strong correlation is observed, reflecting the shared baseline shift.
With correction enabled, the correlation vanishes, indicating that the common-mode contribution has been successfully removed independently in each \gls{CRU}.

These results demonstrate the effectiveness and robustness of the implemented common-mode correction algorithm under high-rate conditions.
\subsubsection{Pedestal subtraction and zero suppression}
\label{sec:UL:Blocks:PedZS}

Each analog channel of the SAMPA chip exhibits an intrinsic electronic offset---known as the pedestal---which is a stable baseline (in the absence of signals) characteristic of the front-end electronics.
This offset biases the sampled \gls{ADC} values and must be subtracted to recover the true signal.
This operation is performed by dedicated pedestal correction cores.
In addition, the common-mode correction is applied to remove the previously calculated baseline shifts shared across multiple channels.
Finally, the signal is compared to a defined channel-specific threshold (zero point), with values below this threshold being suppressed.

Each pedestal correction core processes one link data stream and handles two \gls{ADC} values in parallel.
With \num{20} independent cores operating concurrently, the system achieves the necessary throughput (\num{1600} samples for \num{20} links each 40 clock cycles).

\begin{landscape}
\begin{figure}
  \centering
  \includegraphics[width=0.95\linewidth]{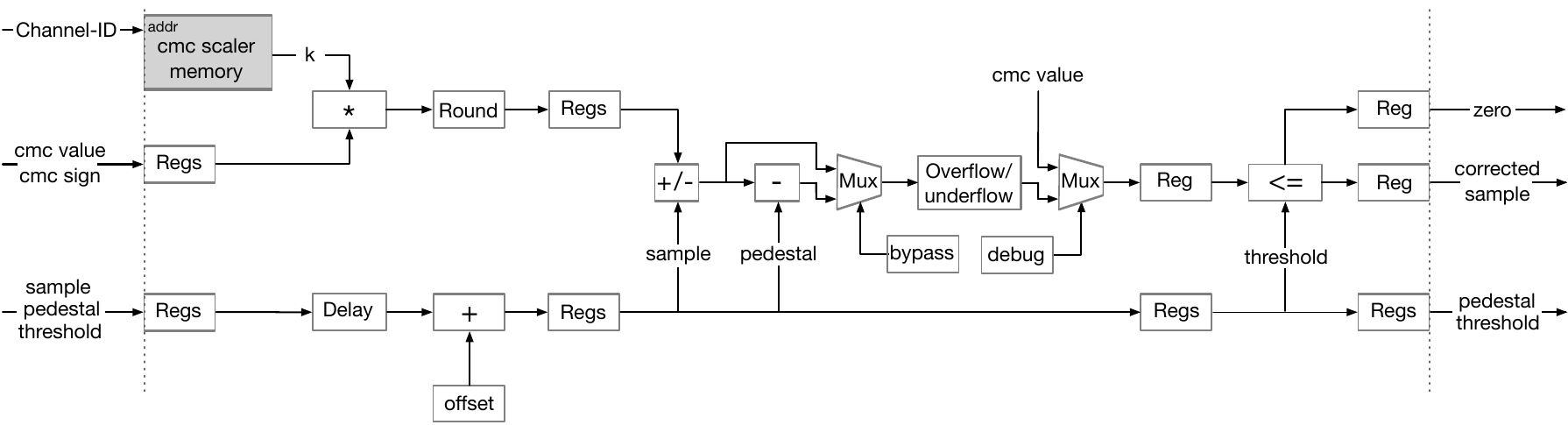}
  \caption{Overview of the pedestal correction core.
  The \textit{Channel-ID} from the time info stream is used to address the lookup of the scaling factor $k_\text{pad}$ for the application of the common-mode correction in the first stage.
  The module outputs the corrected sample and the \textit{Zero} flag used in the link data stream to indicate that the sample is below threshold and may be suppressed.}
  \label{fig:TPC_UL_PEDCORE}
\end{figure} 
\end{landscape}

\Figref{fig:TPC_UL_PEDCORE} shows the processing pipeline.
In the first stage, the channel-specific scaling factor $k_\text{pad}$ for common-mode correction (\cf \secref{sec:UL:Blocks:CMC}) is retrieved from configuration memory.
The \nbit{6} \textit{Channel-ID} from the time info stream is used to address this lookup.
In the subsequent stages, each sample undergoes the common-mode adjustment, pedestal subtraction, and suppression of values below threshold.
Underflow and overflow protection are applied, matching the \gls{ADC}’s dynamic range (\num{0} to \num{1023}).
A debug mode can redirect selected \gls{ADC} channels to output the common-mode value instead of the sample values, aiding diagnostics and calibration.

The pipeline relies on fixed delays to keep all data streams and correction parameters precisely aligned in time, ensuring each sample is processed with its matching common-mode and pedestal values. Consequently, minor delay elements, implemented as registers or shift registers, are inserted as required to ensure proper timing alignment.
\subsubsection{Ion-tail filter}
\label{sec:UL:Blocks:ITF}

The ion-tail filter is a digital signal processing algorithm designed to mitigate characteristic signal distortions caused by slowly drifting ions generated in the induction gap of the \gls{GEM} stack~\cite{CMCITpaper}.
These ions induce an exponentially decaying tail in the pulse shape, which degrades signal quality and, if uncorrected, impacts subsequent data processing and reconstruction.

Two main sources of ion-induced distortion can be identified:
(i) ions generated in the high-field region near the GEM holes, just below the final \gls{GEM} amplification stage—an effect common to all \gls{GEM}-based systems; and
(ii) ions produced throughout the full induction gap as a result of charge amplification along its volume—an effect specific to the high-voltage configuration used in the \gls{ALICE} \gls{TPC}.

\paragraph{Algorithm overview}

The ion-tail filter is implemented as a recursive Infinite Impulse Response (IIR) filter designed to model an exponential decay.
It operates independently on each input channel, iterating over the incoming samples.
At each iteration, the input sample $q_{\text{in}}$ is adjusted using a scaled correction term $q_{\text{cor}}$, yielding the filtered output $q_{\text{out}}$ as:

\begin{equation}
  q_{\text{out}} = q_{\text{in}} - k_1 \cdot (1 - k_2)\cdot q_{\text{cor}} = q_{\text{in}} - k_x\cdot q_{\text{cor}}\ .
  \label{sec:UL:Blocks:itfcorr}
\end{equation}

Concurrently, the correction factor $q_{\text{cor}}$ is updated for use in the subsequent iteration according to the recursive relation:

\begin{equation}
  q_{\text{cor}} = k_2\cdot (q_{\text{in}} + q_{\text{cor}})\ .
  \label{sec:UL:Blocks:itfcorr2}
\end{equation}

The filter behavior---specifically, the degree of smoothing and the strength of the feedback---is governed by the parameters $k_1$ and $k_2$.
To reduce resource usage and avoid redundant computations in the \gls{FPGA} implementation, the coefficient $k_x = k_1\cdot(1 - k_2)$ is precomputed and stored as configuration parameter together with $k_2$.

\begin{figure}
  \centering
  \includegraphics[width=0.99\linewidth]{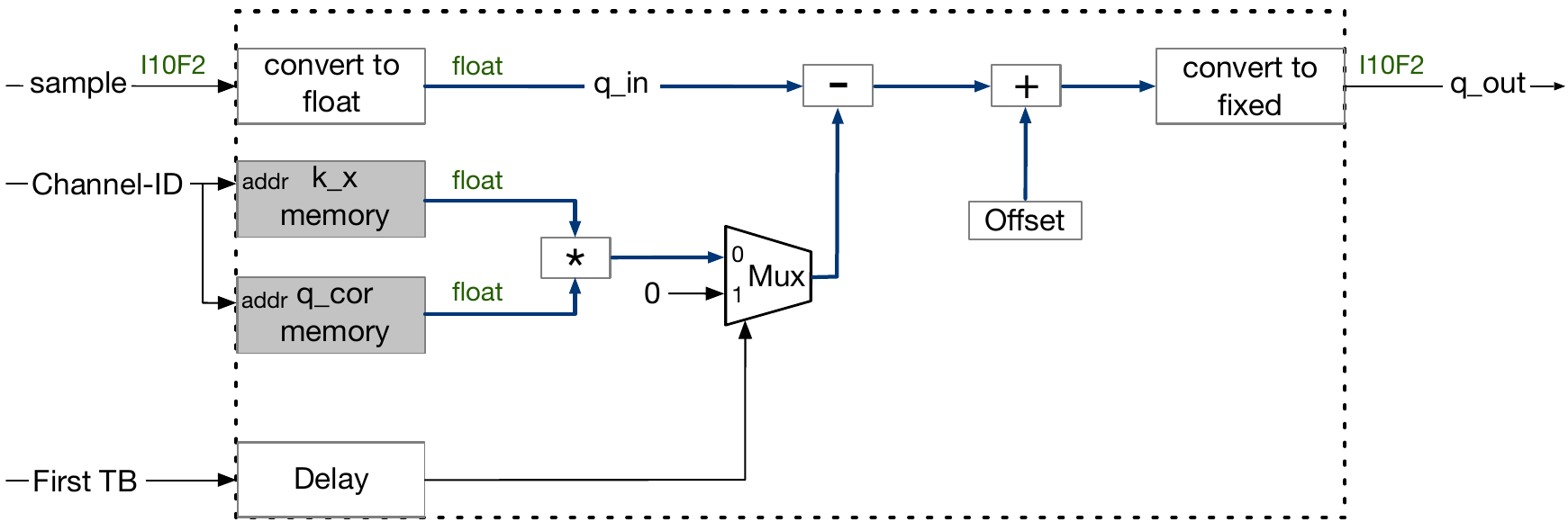}
  \caption{Correction algorithm of an ion-tail filter core.
  The scaling factor $k_x$ and the correction value $q_\text{cor}$ are fetched from memory using the \textit{Channel-ID}.
  Correction of the first sample in a sequence is inhibited using the \textit{First TB} signal, as no valid correction value is available at that point.
  Both signals are available from the time info stream.
  The \textit{First TB} signal is delayed by four clock cycles using a shift register while the sample is converted to single-precision floating-point format, which is used for all computations inside the dashed box.}
  \label{fig:ITF_COR}
\end{figure}

\paragraph{Implementation}

Although the ion-tail correction algorithm is relatively straightforward to implement in both software and firmware, the primary challenge arises from the high channel count (up to \num{1600} channels) and the stringent timing constraint of \num{48} clock cycles.
This requirement necessitates the simultaneous processing of \num{40} channels in parallel.
Additionally, higher numerical precision than the native \nbit{12} fixed-point format is desirable to maintain computational accuracy throughout the correction process.
To address this, all \nbit{12} input samples are first converted to \nbit{32} floating-point representations.
All arithmetic operations are performed in floating-point, and the final result is subsequently converted back to a \nbit{12} fixed-point integer format.

\subparagraph{Ion-tail filter core}

The ion-tail filter is implemented using \num{40} dedicated filter cores, each operating in parallel.
Two cores are assigned to each link data stream:
one handles even-numbered channels, and the other processes odd-numbered channels, with each core managing \num{40} channels concurrently.
The correction logic of an individual core is depicted in \figref{fig:ITF_COR}.
Upon receiving a sample, the core converts it into a floating-point value.
Simultaneously, the \textit{Channel-ID} from the time info stream is used to retrieve the correction factor $q_\text{cor}$ and scaling factor $k_x$ for the given channel from memory.
The scaled correction is computed as the product of the correction and scaling values and is then subtracted from the floating-point representation of the input sample, following \eqref{sec:UL:Blocks:itfcorr}.

For testing and debugging purposes, a configurable offset can be applied after the correction and before converting the result back to fixed-point format.
In the case of the first sample in a sequence, where no valid correction factor exists, the scaled correction term can be effectively bypassed by forcing its value to zero via a multiplexer connected to the delay unit.

The processing pipeline is fully unrolled into \num{13} sequential stages without internal feedback loops, enabling a throughput of one sample per clock cycle and a fixed latency of \num{13} clock cycles.

\subparagraph{Correction factor update}

\begin{figure}
  \centering
  \includegraphics[width=0.9\linewidth]{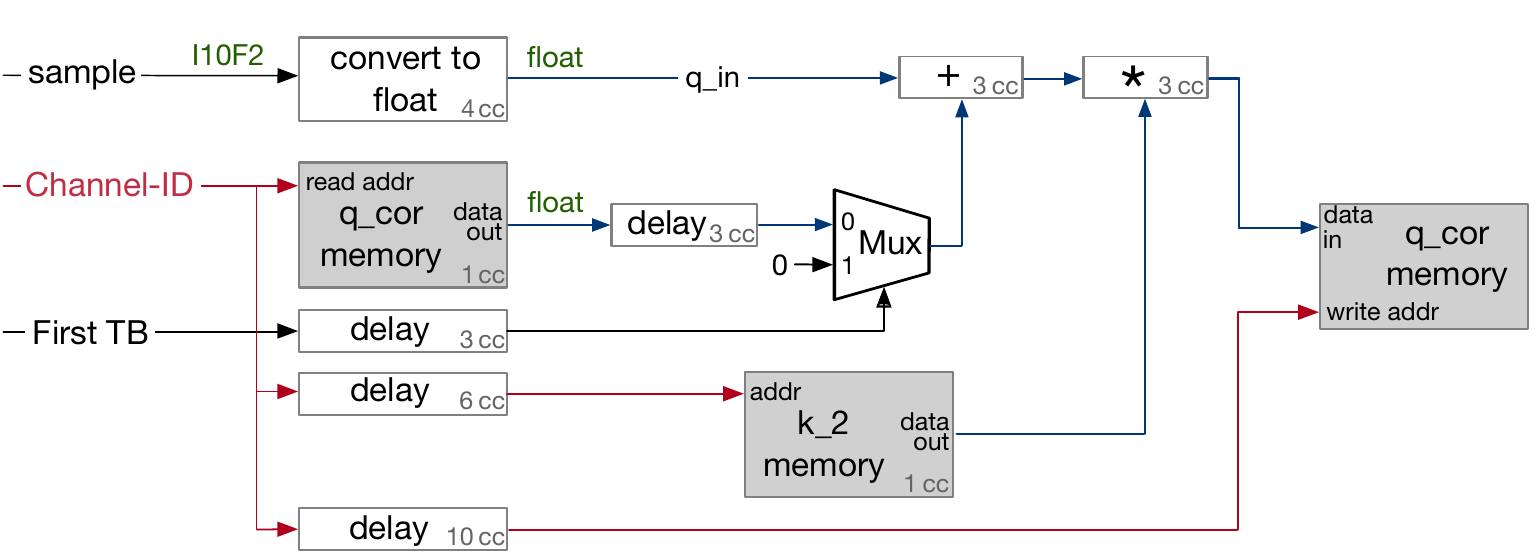}
  \caption{Dual-port RAM is used to store the correction factor $q_{\text{cor}}$.
  The chosen architecture supports concurrent read and write operations, enabling iterative updates of $q_{\text{cor}}$.
  The coefficients $q_{\text{cor}}$ and $k_2$ are retrieved from memory using the \textit{Channel-ID}.
  The value of $q_{\text{cor}}$ is forced to zero for the first time-bin in a sequence.  
  The sample is converted to single-precision floating-point format to improve numerical accuracy in the subsequent calculations.
  The diagram also indicates the latencies of the individual operations in clock cycles (cc).}
  \label{fig:ITF_COR_UPDATE}
\end{figure}

\Figref{fig:ITF_COR_UPDATE} illustrates the logic responsible for the iterative update of the correction factor $q_{\text{cor}}$ following \eqref{sec:UL:Blocks:itfcorr2}.
At the core of the design is a dual-port memory, which stores the current values of $q_{\text{cor}}$ for all 40 channels.
This memory architecture enables concurrent read and write operations through dedicated ports, facilitating efficient real-time updates.

During each processing cycle, the input sample is converted to a \nbit{32} floating-point value $q_{\text{in}}$, while the corresponding $q_{\text{cor}}$ value is simultaneously retrieved from memory.
To compensate for the latency introduced by the floating-point conversion, the $q_{\text{cor}}$ value is delayed by three clock cycles.
The updated correction value is then computed by summing $q_{\text{in}}$ and $q_{\text{cor}}$, followed by multiplication with the filter coefficient $k_2$, following \eqref{sec:UL:Blocks:itfcorr2}.
The resulting value is subsequently written back to memory via the write port.

The update is not executed when \textit{ADC valid} in the time info stream indicates that no valid sample data is present in the current clock cycle (\cf~\secref{sec:UL:Dataformat}).
This is not shown in \figref{fig:ITF_COR_UPDATE}.

The update logic is highly pipelined and optimized for resource efficiency.
Rather than delaying the \nbit{32} coefficient $k_2$ through each pipeline stage, the design instead delays the \nbit{6} channel address, significantly reducing register usage.
This approach yields substantial resource savings.%
\footnote{The savings amount to $(32 - 6)$ bits\ $\times\ 6$ pipeline stages\ $\times\ 40$ cores\ $= 6240$ registers ($0.5$\,\% of all available Flip-Flops).}
This optimization is applicable only to $k_2$; the $q_{\text{cor}}$ values themselves cannot be delayed in this manner, as they are also required without latency in the correction stage.

Further resource savings are achieved by utilizing small, embedded memory blocks to store $q_{\text{cor}}$ values, rather than conventional register arrays.
A register-based implementation would require \num{51200} registers%
\footnote{$1600$ channels\ $\times 32$ bits\ $= 51200$ registers  ($4$\,\% of all available Flip-Flops).}
in addition to logic for address decoding and control.
By leveraging embedded memory, these overheads are eliminated, enabling a compact and efficient hardware implementation.

\paragraph{Performance}

The performance of the ion-tail filter is best evaluated using dedicated laser calibration data.
Laser-induced signals have well-defined amplitudes and highly reproducible pulse shapes, which makes them particularly suitable for studying the ion tail and its correction.
This allows a detailed statistical characterization of the tail structure and the optimisation of the filter parameters.
A comprehensive description of this procedure and its quantitative evaluation is given in \cite{CMCITpaper}.
With physics data, the filter manifests itself primarily as a reduction of the recorded data size of about one percent, consistent with the suppression of residual signal tails.
\subsubsection{Integrated digital currents}
\label{sec:UL:IDC}

The \glspl{IDC} are cumulative sums of raw \gls{ADC} samples from each channel over a defined time window.
They are used in calibration procedures to correct distortions caused by space charge in the drift volume.
Within the \gls{UL}, the \gls{IDC} processor performs this integration in parallel with the main raw-data processing pipeline (\cf \figref{fig:UL:Overview}).

\begin{figure}[ht]
  \centering
  \includegraphics[width=0.8\linewidth]{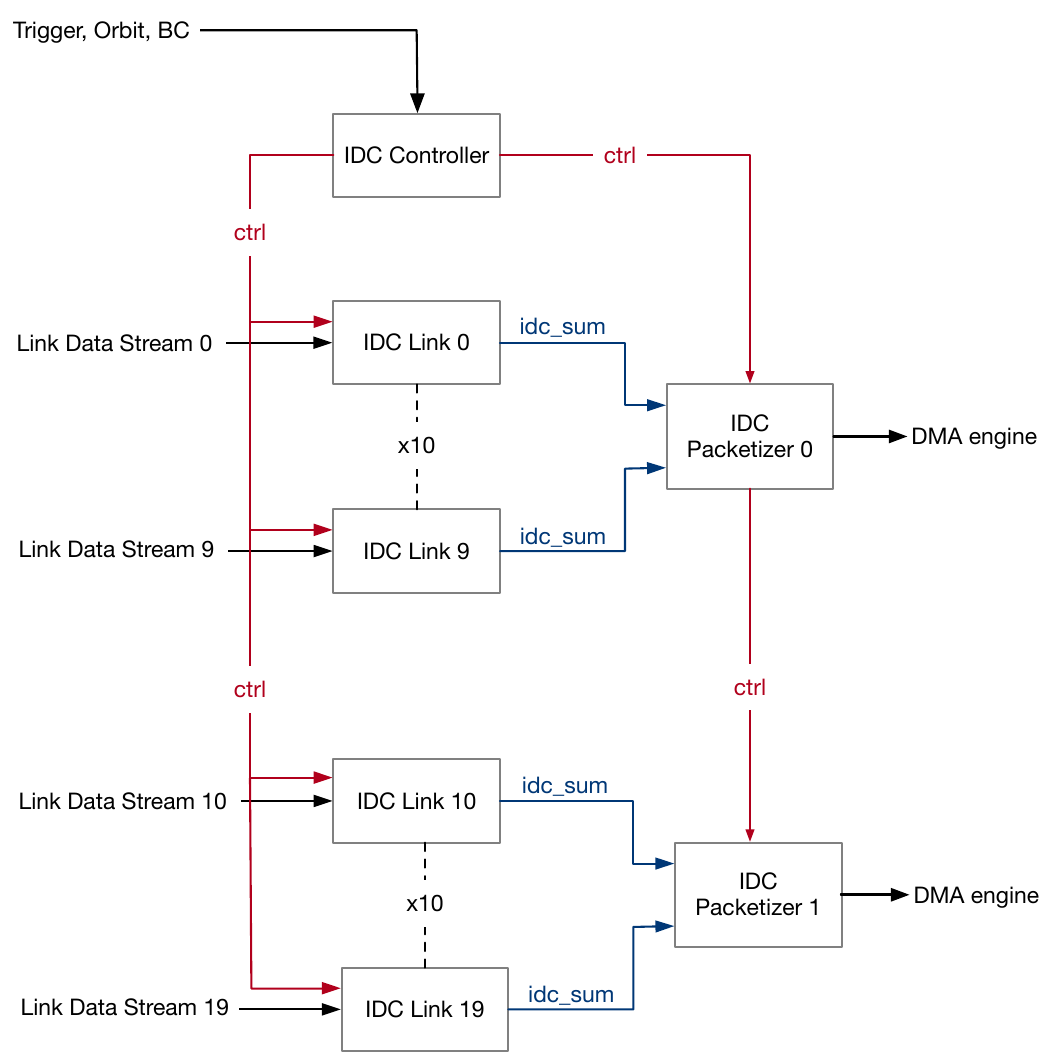}
  \caption{IDC processor.
  The link cores integrate incoming samples, while the packetizer formats the results into packets.
  All operations are coordinated by the controller.}
  \label{fig:TPC_UL_IDC_PROC}
\end{figure} 

As illustrated in \figref{fig:TPC_UL_IDC_PROC}, the \gls{IDC} processor consists of three main components: the \textit{link core}, the \textit{controller}, and the \textit{packetizer}.
The link cores perform the per-channel integrations; \num{20} cores operate in parallel to cover up to \num{1600} channels, each producing \num{80} sums.
These values are forwarded to the packetizer, which formats them into packets for Direct Memory Access (DMA) transfer.

The controller coordinates the operation of the processor.
It defines the integration windows using external timing signals such as orbit, bunch-crossing (BC), or trigger inputs, configures the link cores, and initiates packet transmission once the accumulated data are ready.

The \gls{IDC} processor uses a dedicated packet format and transmission channel, allowing the resulting data to be processed asynchronously by the online farm without interfering with the main data stream.

\begin{figure}
  \centering
  \includegraphics[width=0.7\linewidth]{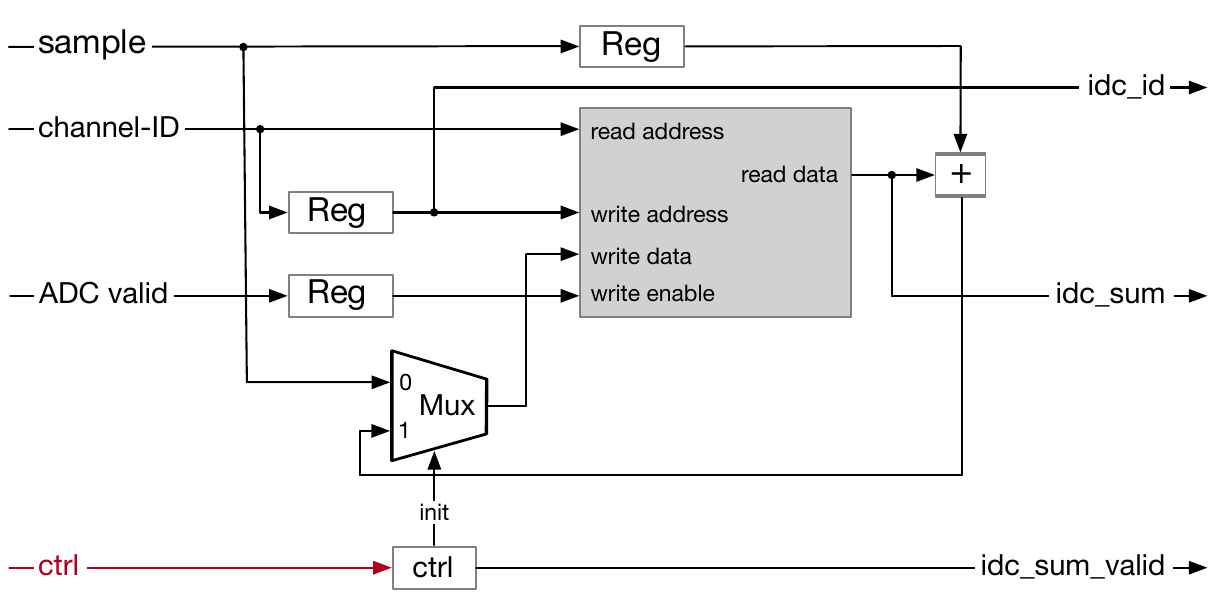}
  \caption{The IDC link core.
  A dual-port memory stores the accumulated sums.
  New samples are added to the stored values and written back in the next cycle.}
  \label{fig:TPC_UL_IDC_LINK}
\end{figure}

The link core is the most resource-intensive component, as \num{20} instances operate in parallel and each maintains sums for \num{80} channels.
Its structure is shown in \figref{fig:TPC_UL_IDC_LINK}.
The core uses a small dual-port memory to store the accumulated sums.
During integration, each incoming sample is added to the corresponding stored value, and the updated sum is written back in the next cycle.

At the end of an integration window, the accumulated values are read out sequentially while new samples continue to be integrated.
This interleaved scheme allows continuous operation without interrupting the data flow.

\paragraph{Performance}

The \glspl{IDC} are a key input for calibrating space-charge distortions in the \gls{TPC}.
At high interaction rates, ions produced in the amplification stage accumulate in the drift volume, distorting the electric field and affecting reconstructed track positions.
Local variations of the space-charge density are observed both in time, due to fluctuations in collision rate and charged-particle multiplicity, and in space, reflecting the non-uniform distribution of particle tracks across the detector.
The \glspl{IDC} provide a time- and region-resolved measurement of the charge collected in different pad areas, enabling correction of the space-charge fluctuations on millisecond time scales and restoring the spatial accuracy of the recorded data.
A detailed analysis of this performance is beyond the scope of this paper.

\subsection{Output stage}
\label{sec:UL:Packing}

The output stage receives the link data streams and the time info stream, extracts the corrected samples, and packs them into a dense, zero-suppressed format.
The goal is to generate compact data blocks with minimal padding and size-optimized headers, with the aim to significantly reduce the total \gls{TPC} output data volume, keeping it well below \SI{1}{\tera\byte\per\second}, as required by the bandwidth constraints of the downstream processing stages.

\gls{ALICE}'s readout framework mandates that the detector data corresponding to a single \gls{HBF} be encapsulated into one or more contiguous packets, each with a maximum size of \SI{8}{\kibi\byte}.
To meet this requirement, each of the two dense packing instances (\cf~\figref{fig:UL:Overview}) is capable of producing correctly formatted packets from up to \num{10} detector links operating at full data rate.
The packet format has been optimized for efficient parallel processing on \glspl{GPU} within the \gls{EPN} farm.

\subsubsection{Data format}
\label{sec:UL:DataFormat}

The produced packets conform to \gls{ALICE} readout requirements, specifically: each packet is at most \SI{8}{\kiB} in size and begins with a \gls{RDH}~\cite{AliceReadout, CruFirmware}.
The payload is structured as a sequence of time-bin blocks.
Each block begins with a byte-aligned block header that includes the number of contributing links and the bunch-crossing identifier. Following the block header, link headers are included for each active link.
These headers are then followed by the data payloads corresponding to each active link.

Per detector link and time-bin, up to \num{80} extended \nbit{12} samples can arrive at the input of the Dense Packing unit.
To efficiently identify which samples are included in a payload, the link headers implement a compact two-level bitmask scheme.%
\footnote{The time-bin block header has a fixed size of \SI{2}{\byte}.
Link header sizes range between \SI{3}{\byte} and \SI{10}{\byte}, depending on the number of active channels.
The maximum size of a link data payload is $\num{80} \times \SI{12}{\bit} = \SI{120}{\byte}$.
Thus, the maximum size of a time-bin block is $\SI{2}{\byte} + 10 \times \SI{10}{\byte} + 10 \times \SI{120}{\byte} = \SI{1302}{\byte}$.}
The scheme incorporates a header bit that specifies whether the bit mask is static or dynamic.
In the static configuration, the bit mask is fixed and comprises 80\,bits, each representing the presence or absence of a corresponding sample.
In the dynamic configuration, the \num{80} samples are partitioned into ten groups of eight samples, and a \nbit{10} mask is used to indicate the presence or absence of each group.

Zero-valued samples are suppressed, and the remaining non-zero \nbit{12} samples are densely packed without regard to byte boundaries.%
\footnote{Only at the end of a time-bin block may a \nbit{4} padding be added to ensure byte alignment for the next block.}
Each packet may contain zero or more partial or complete time-bin blocks.
To minimize packet overhead, packets are filled to the maximum size whenever possible.
Consequently, a single time-bin block may be split across consecutive packets within a \gls{HBF} when approaching the \SI{8}{\kiB} size limit.
Only the first packet of a \gls{HBF} is guaranteed to begin with a full time-bin block.

Depending on the channel activity, the number of packets per \gls{HBF} varies from as few as one (in the case of no activity) up to \num{72} for full-channel activity.
Each packet ends with a \nbit{128} meta word that contains metadata including the number of time-bins in the packet, the number of samples, and the offset of the first time-bin block within the packet.
This meta word enables precise access to time-bins within each packet and facilitates the correct time-wise alignment and distribution of data across different \glspl{STF} to the \glspl{GPU} for reconstruction.

Additionally, the final packet of each \gls{HBF} includes a \nbit{128} trigger word positioned immediately before the meta word. This word provides information about the types and timing of all triggers received during the \gls{HBF}.

\subsubsection{Merging and packing pipeline}
\label{sec:UL:MergingPacking}

\begin{landscape}
\begin{figure}
    \centering
    \includegraphics[width=\linewidth]{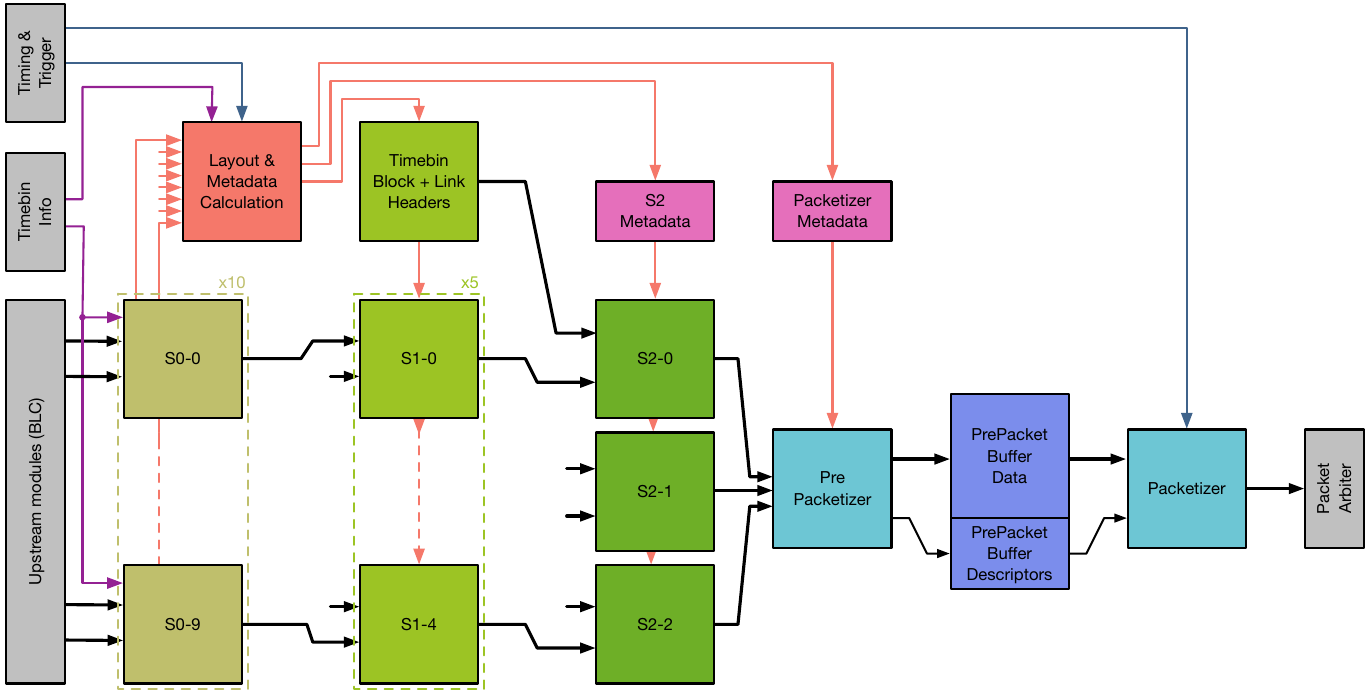}
    \caption{Simplified overview of the pipeline packing the data of up to \num{10} link data streams into one or more densely packed packets of up to \SI{8}{\kiB}.}
    \label{fig:UL:Packing:overview}
\end{figure}
\end{landscape}

\Figref{fig:UL:Packing:overview} illustrates the architecture of the dense packing pipeline.
The pipeline's first three stages operate on a time-bin basis and progressively merge data into wider data words until the target output width of \SI{256}{\bit} is achieved.
The later stages wrap the resulting stream of time-bins with an \gls{RDH}, along with meta and trigger words, to form \SI{8}{\kiB} packets ready for transmission.
The operation of all pipeline components is coordinated by a central layout calculation unit.

At \stage{0}, ten parallel instances---one per link data stream---pack the incoming corrected \nbit{12} samples into dense \nbit{64} words.
These words are written to FIFOs that serve as inter-stage buffers.
Depending on the number of non-zero samples in a given time-bin, each \stage{0} unit generates between \num{1} and \num{15} words.
All words, except potentially the last, are fully populated with samples; the final word may be zero-padded.

Once a \stage{0} unit finishes processing a time-bin, the number of non-zero samples per link is known.
Based on this and the end position of the previous time-bin, the layout calculation unit determines the sizes and start positions for each link’s payload and header in the current time-bin block.
It also accumulates metadata for the meta word and maintains overall packet layout throughout the \gls{HBF}, ensuring correct placement of \gls{RDH}, placeholder regions, and final metadata.

The data merging scheme relies on computing the current position and contribution size for each time-bin, and precalculating the target positions for subsequent stages.
Stream widths expand from \SI{64}{bit} in \stage{0} to \SI{128}{bit} in \stage{1} and \SI{256}{bit} in \stage{2}.
As merging is performed in parallel, each stream is shifted to its designated position and zero-padded at both ends, enabling seamless concatenation by applying a logical \textit{OR} between adjacent streams.

\stage{1} comprises one header generation unit and five regular merging units.
The header unit produces the time-bin block header and link headers, while each merger processes data from two \stage{0} instances.

All \stage{1} units follow a common operational scheme:
The unit controllers use barrel shifters to align incoming \nbit{64} data words to the positions specified by the layout unit.
These aligned words are written into asymmetric FIFOs with \nbit{64} input and \nbit{128} output width, acting as buffers toward \stage{2}.
To maintain correct alignment at the \nbit{128} level, zero-padding may be inserted before the first and after the last data words.

At \stage{2}, the pipeline merges prealigned and padded \nbit{128} words into wider \nbit{256} words using asymmetric FIFOs (\nbit{128} in\,/\,\nbit{256} out).
These units also insert placeholders for the meta and trigger words at predefined positions.
Thanks to the alignment information provided by the layout unit, all \stage{1} and \stage{2} components can operate in parallel and complete the packing of a single time-bin within the available \num{48} clock cycles at \SI{240}{\mega\hertz}.

The prepacketizer merges the output from the three \stage{2} units and replaces the previously inserted placeholders with actual meta and trigger data collected by the layout unit.
Whereas the earlier stages operate on time-bins, the prepacketizer outputs prepackets---complete packet structures except for the \gls{RDH}---along with a prepacket descriptor.
This descriptor is then passed to the packetizer, which uses its contents to generate the final \gls{RDH}.
It then assembles and transmits full packets in \nbit{256} words to the Common Logic, which handles forwarding to the \gls{FLP}.

\section{User Logic for auxiliary systems}
\label{sec:ULSync}

\begin{figure}[ht]
  \centering
  \includegraphics[width=0.6\linewidth]{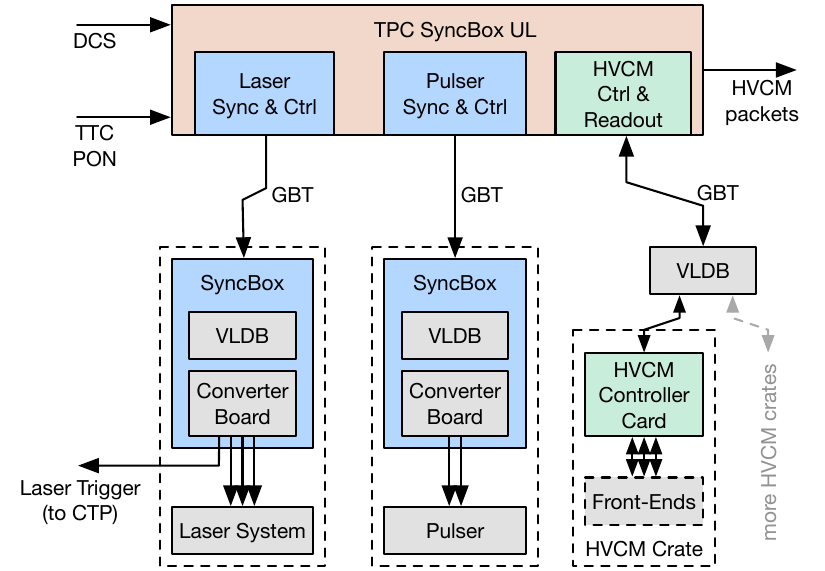}
  \caption{Building blocks of the Synchronisation Box (SyncBox) \gls{UL} and interfaces to the external systems.}
  \label{fig:ULSync:overview}
\end{figure}

The \gls{TPC} includes several auxiliary systems designed to provide stimuli and additional data to ensure correct operation and enhance calibration.
A key requirement for these systems is that their operation must be synchronized with the sampling clock of the SAMPAs and the TPC data readout.
Given its interface with the experiment’s trigger and timing system, and its use of the \gls{GBT} communication channel with fixed latency, the \gls{CRU} is well suited to implement the necessary control and readout logic for these auxiliary systems.
\Figref{fig:ULSync:overview} presents a simplified overview of the \gls{SyncBox} \gls{UL}, which is responsible for controlling the laser and pulser systems and managing the readout of the \gls{HVCM} systems.

To enable control of the laser and pulser systems via the \gls{GBT}, several \gls{SyncBox}es were developed.
Each \gls{SyncBox} consists of a \gls{VLDB}~\cite{vldb}, equipped with a \gls{VTRx} and \gls{GBTx}, and a custom-designed interface board.
The \gls{VLDB} forwards the \gls{GBT} downlink signals received via its \gls{ELink} to the interface board with deterministic latency.
There, the signals are converted into optical and electrical outputs at the required levels.
All logic required to generate properly timed and synchronized control signals is implemented within the \gls{CRU}.

\subsection{Laser control}
\label{sec:ULSync:laser}

Pulsed UV lasers~\cite{lasers} are used to generate reference tracks in the active volume of the \gls{TPC}, supporting alignment, calibration, and drift-time monitoring.

The converter board provides five output signals\footnote{For Q-switch, flash, shutter, and camera synchronization, and a spare channel.} for each of the two lasers.
These outputs are controlled by two independent laser controller instances implemented in the \gls{SyncBox} \gls{UL}.

For reliable data analysis, laser firing must be synchronized with the SAMPA sampling clock and aligned with the \glspl{TF} (\cf \secref{sec:TPC:overview}), ensuring that laser signals appear at well-defined positions in the data stream.

The controller generates a periodic start signal defining time windows during which a laser sequence may be initiated.
Its frequency is approximately \SI{10}{\Hz}, corresponding to the fixed repetition rate of the lasers.
The associated counter runs synchronously with the \gls{LHC} clock and is reset by the same trigger used to synchronize the SAMPAs.
The start signal frequency is chosen as an even integer multiple of the \gls{TF} frequency, ensuring synchronization with both the sampling clock and the readout.

Laser events are embedded in regular collision data taking, allowing monitoring and correction of drift-velocity variations in space and time.
To facilitate their identification, the laser controller sends a trigger to the \gls{CTP}, enabling the \glspl{CRU} to tag data packets containing laser events.

The timing and duration of each output signal are configurable relative to the start signal, allowing the laser pulses to be placed safely within a \gls{TF} and sufficiently far from its boundaries.

\subsection{Calibration pulser control}
\label{sec:ULSync:pulser}

The \gls{TPC} calibration pulser system injects charge onto the readout pads, allowing the full electronics chain to be tested with a well-defined input signal.
One pulser channel is provided per \gls{GEM} stack, corresponding to four channels per sector and 144 channels for the entire \gls{TPC}.
The electronics is distributed over two crates with 72 channels each, one for each side of the detector.

Similar to the laser control, the \gls{UL} provides a \SI{5}{\MHz} pulser clock synchronized with the SAMPA sampling clock.
When a pulser firing is requested—typically via the trigger system—the controller generates a signal aligned with this clock and adapted to the pulser interface.
Together with the pulser clock, the signal is transmitted to the pulser via the corresponding pulser \gls{SyncBox}.

\subsection{Readout of the high voltage current monitor system}
\label{sec:ULSync:HVCM}

The \gls{HVCM} provides dynamic estimates of the space charge generated by backflowing ions by measuring the currents at the top electrode of the lowest \gls{GEM} foil (G4T) in each \gls{GEM} stack.
The currents are sampled at \SI{1}{\kHz} with a resolution of \SI{1}{\nano\A}.%
\footnote{The resolution is dominated by the capacitance of the \SI{60}{m} long HV cables.
The intrinsic resolution of the \gls{HVCM} electronics has been measured to be \SI{3}{\pico\A}.}
For the \gls{TPC}, two \gls{HVCM} crates are employed.

To make the current measurements immediately available and correlate them with detector data, the readout and timestamp synchronization are implemented in the \gls{UL} of the \gls{CRU}.
Each crate contains nine front-end cards with eight measurement channels each.
The front-end cards transmit formatted data streams containing timestamps and current measurements via \gls{UART} to a central controller card.
The controller aggregates the data into \SI{1}{\kiB} frames, each containing payload from a single front-end card, and transmits them in a round-robin sequence using \ebtb encoding over an \gls{ELink} to a \gls{VLDB}.

An additional crate with the same data format is used to read currents from the central electrode and field cage.
Although the resulting data rates are modest, all packets must be preserved in order to reconstruct the complete current history.

To allow precise temporal correlation with detector data, the timestamps of all front-end streams are reset synchronously.
At the beginning of a run, the \gls{UL} distributes a trigger signal via the latency-deterministic \gls{GBT} downlink to all connected \gls{HVCM} crates, where it is internally propagated to reset the timestamps of all front-end cards simultaneously.

\begin{figure}[h]
  \centering
  \includegraphics[width=.99\linewidth]{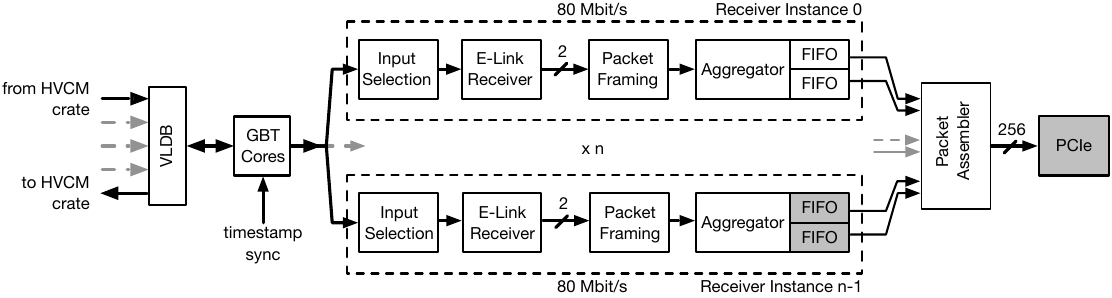}
  \caption{High-voltage current monitor readout chain.}
  \label{fig:ULSync:hvcmreadout}
\end{figure}

The readout chain is illustrated in \figref{fig:ULSync:hvcmreadout}.
Monitoring data from the crates are transmitted via the \gls{GBT} to the \gls{CRU} at \SI{80}{\mega\bit\per\second} per stream.
Dedicated receiver instances perform deserialization, decoding, and frame reconstruction.
Synchronization is achieved using \ebtb idle symbols, which appear between frames.
Each reconstructed frame is validated using its \gls{CRC}, and a corresponding frame descriptor is generated.
Frames and descriptors are buffered in FIFOs before packet assembly.

The packet assembler creates readout packets consisting of an \gls{RDH}, followed by a \nbit{256} \gls{CRU} header containing sequence numbers, timing metadata, and error flags.
The packet is completed by appending the \SI{1}{\kiB} \gls{HVCM} data frame.

\section{Summary and discussion}

Since the start of \gls{LHC} operation in 2009, \glspl{FPGA} have been a central component of data acquisition and online processing in the major experiments.
Their reconfigurable logic, high I/O density, and deterministic timing make them ideally suited to handle the enormous bandwidths and low-latency requirements of modern detectors.

Over the last fifteen years, their role has shifted from basic data formatting to increasingly physics-aware online processing.
ATLAS introduced FPGA-based calorimeter Feature Extractors to produce trigger primitives directly from finely segmented inputs~\cite{AtlasFex}.
LHCb adopted the TELL1 boards as its principal readout platform~\cite{LhcbTell1}, where the \glspl{FPGA} executed real-time preprocessing such as pedestal subtraction, zero suppression, and hit clustering.
\gls{ALICE} integrated large \gls{FPGA} farms into its \gls{HLT} system, enabling online cluster finding, tracking, and calibration~\cite{AliceCrorc}.
By Run 3, ATLAS \cite{AltalsFelix}, LHCb \cite{LHCbupgrade}, and ALICE \cite{TDR:rdo} had all transitioned to PCIe-based FPGA readout boards—including the widely adopted PCIe40 platform \cite{pcie40}---but with firmware stacks tailored to their specific detector architectures and processing requirements.

\gls{ALICE}, in particular, has consistently pushed \gls{FPGA}-based readout a step further.
In Run\,1, FPGA cluster finding enabled fast online processing in the \gls{HLT} farm, while in Run\,2, the \gls{HLT} discarded raw data and stored only clusters and tracks~\cite{AliceHlt}.
In Run\,3, the introduction of \glspl{CRU} has extended the role of off-detector \glspl{FPGA} into a full-fledged data conditioning and preprocessing layer between the detector front-ends and the \gls{GPU}-based \glspl{EPN}.

The approach adopted for the \gls{ALICE} \gls{TPC} in Run,3, described in this paper, emphasizes \textit{simplicity}, \textit{robustness}, and \textit{flexibility}.
On-detector logic was deliberately minimized: raw digitized signals are transmitted over optical links based on the radiation-tolerant \gls{GBT} chipset and Versatile Link components, shifting the processing burden to off-detector \glspl{FPGA}, where algorithms can be more easily updated and refined as operational experience accumulates.
Reliability was a guiding principle throughout; for example, the SAMPA chip can recover immediately after reset, restoring the full \gls{TPC} \gls{FEE} in less than \SI{400}{\ns}.

Radiation-induced link losses were nevertheless observed during operation and were traced to a voltage regulator on the \gls{FEC}.
These losses disrupt data alignment at the \gls{CRU} for the affected link.
However, periodic re-synchronization of the detector restores proper alignment, so that in practice the overall operation and performance of the readout remain unaffected.

Together with this mitigation strategy, the chosen approach has resulted in very stable operation, with the \gls{TPC} readout functioning reliably under all running conditions, both with and without beam.

The \gls{TPCUL} firmware described in this paper implements a continuous, triggerless readout for the \gls{ALICE} \gls{TPC} at interaction rates up to \SI{50}{\kHz}.
At a sustained real-time processing rate of \SI{3.3}{\tera\byte\per\second}, \num{1600} new \gls{ADC} samples arrive at each readout card every \SI{200}{\ns}, requiring all processing stages to operate deterministically within fixed latency constraints.
The processing chain includes algorithms essential for \gls{TPC} operation and performance optimization:
(1) common-mode correction across up to 1600 channels per \gls{CRU},
(2) ion-tail filtering and baseline correction,
(3) zero suppression, and
(4) dense data packing optimized for subsequent \gls{GPU} processing.
These functions provide the necessary data reduction and conditioning for high-throughput continuous readout, while carefully optimizing \gls{FPGA} resource usage.
Extensive validation in simulation and with beam data demonstrates reliable performance of the common-mode correction and other preprocessing steps.
The firmware also supports online calibration by computing integrated digital currents.

Beyond data processing, the firmware integrates control of auxiliary subsystems, such as laser and pulser operation and high-voltage current monitor readout, consolidating functions to reduce heterogeneity and simplify maintenance.

In summary, the \gls{ALICE} \gls{TPC} readout combines lightweight on-detector electronics, radiation-tolerant optical links, and flexible off-detector \gls{FPGA} processing into a robust solution.
It meets the demands of unprecedented cluster and track densities at \SI{50}{\kHz} \PbPb collisions, applies the necessary corrections to ensure high data quality, and has proven very stable in operation.

\section*{Acknowledgements}

We gratefully acknowledge the central \gls{CRU} team for their support, expertise, and valuable discussions.
We extend special thanks to Filippo Costa for his dedicated assistance, commitment, and prompt responses to all our questions and requests.
His contributions have been invaluable to this work.

Furthermore, we acknowledges the following funding agencies for their support in the \gls{TPC} Upgrade:
Funda\c{c}\~ao de Amparo \`a Pesquisa do Estado de S\~ao Paulo (FAPESP), Brasil;
Ministry of Science and Education, Croatia;
The Danish Council for Independent Research | Natural Sciences, the Carlsberg Foundation and Danish National Research Foundation (DNRF), Denmark;
Bundesministerium f\"{u}r Bildung, Wissenschaft, Forschung und Technologie (BMBF), 
GSI Helmholtzzentrum f\"{u}r Schwerionenforschung GmbH, 
DFG Cluster of Excellence "Origin and Structure of the Universe", 
The Helmholtz International Center for FAIR (HIC for FAIR)
and the ExtreMe Matter Institute EMMI at the GSI Helmholtzzentrum f\"{u}r Schwerionenforschung, Germany;
National Research, Development and Innovation Office, Hungary;
Nagasaki Institute of Applied Science (IIST)
and the University of Tokyo, Japan;
Fondo de Cooperaci\'{o}n Internacional en Ciencia y Technolog\'{i}­a (FONCICYT), Mexico;
The Research Council of Norway, Norway; Ministry of Science and Higher Education and National Science Centre, Poland;
Ministry of Education and Scientific Research, Institute of Atomic Physics and Ministry of Research and Innovation, and Institute of Atomic Physics, Romania;
Ministry of Education, Science, Research and Sport of the Slovak Republic, Slovakia;
Swedish Research Council (VR), Sweden;
United States Department of Energy, Office of Nuclear Physics (DOE NP), United States of America.\\[2ex]

\bibliographystyle{JHEP_mod}
\bibliography{bibliography.bib}

\end{document}